\documentclass[aps,prx,superscriptaddress,twocolumn,showpacs,longbibliography,nofootinbib]{revtex4-1}
\usepackage{amssymb}
\usepackage{amsmath}
\usepackage{graphicx}
\usepackage[caption=false]{subfig}
\usepackage{amsfonts}
\usepackage{hyperref}
\usepackage{color}
\usepackage{comment}
\usepackage{soul}

\usepackage[normalem]{ulem}

\newcommand{\bea}{\begin{eqnarray}}
\newcommand{\eea}{\end{eqnarray}}
\newcommand{\myskip}[1]{}
\newcommand{\comments}[1]{}

\newcommand{\ee}{\mathrm{e}}
\newcommand{\ii}{\mathrm{i}}
\newcommand{\dd}{\mathrm{d}}

\newcommand{\bigo}[1]{\mathcal{O}\left (#1\right)}

\newcommand{\rom}[1]{\uppercase\expandafter{\romannumeral #1\relax}}
\newcommand{\norbra}[1]{\left( #1\right)}
\newcommand{\sqrbra}[1]{\left[ #1\right]}

\newcommand{\J}{\mathbb{J}}
\newcommand{\id}{\mathbb{I}}

\newcommand{\Tr}{\mathrm{Tr}} 
\newcommand{\average}[1]{\left\langle #1\right\rangle}
  \newcommand{\Wdis}{W_{\rm diss}}

\newcommand{\sys}[1]{#1_{\scriptscriptstyle (S)}}
\newcommand{\bath}[1]{#1^{\scriptscriptstyle (B)}}

\usepackage{hyperref}
\begin{document}
\title{Speed-ups to isothermality: Enhanced quantum thermal machines through control of the system-bath coupling
}

\author{Nicola Pancotti}
\affiliation{Max-Planck-Institut f\"ur Quantenoptik, D-85748 Garching, Germany}
\affiliation{Department of Physics, Technische Universit\"at M\"unchen, 85748 Garching, Germany}
\author{Matteo Scandi}
\affiliation{ICFO - Institut de Ciencies Fotoniques, The Barcelona Institute of
Science and Technology, Castelldefels (Barcelona), 08860, Spain}
\author{Mark T. Mitchison}
\affiliation{School of Physics, Trinity College Dublin, College Green, Dublin 2, Ireland}
\author{Mart\' i Perarnau-Llobet}
\affiliation{Max-Planck-Institut f\"ur Quantenoptik, D-85748 Garching, Germany}

\begin{abstract}
Isothermal transformations are minimally dissipative but slow processes, as the system needs to remain close to thermal equilibrium along the protocol. Here, we show that smoothly modifying the system-bath interaction can significantly speed up such transformations. In particular, we construct protocols where the overall dissipation $W_{\rm diss}$ decays with the total time $\tau_{\rm tot}$ of the protocol as  $W_{\rm diss} \propto \tau_{\rm tot}^{-2\alpha-1}$, where each value $\alpha > 0$ can be obtained by a suitable modification of the interaction, whereas $\alpha=0$ corresponds to a standard isothermal process  where the system-bath interaction remains constant. Considering heat engines based on such speed-ups, we show that the corresponding efficiency at maximum power interpolates between the Curzon-Ahlborn efficiency for $\alpha =0$ and  the Carnot efficiency for $\alpha \to \infty$.  Analogous enhancements are obtained for the coefficient of performance  of refrigerators. We confirm our analytical results with two numerical examples where $\alpha = 1/2$, namely the time-dependent Caldeira-Leggett and resonant-level models, with strong system-environment correlations taken fully into account. We highlight the possibility of implementing our proposed speed-ups with ultracold atomic impurities and mesoscopic electronic devices.
 \end{abstract}
\maketitle

\section{Introduction}

Isothermal transformations play a fundamental role in thermodynamics, being the building block of optimal processes such as the Carnot engine~\cite{callen1998thermodynamics}. In principle, however, they are infinitesimally slow;  this means in practice that the total time of the process needs to be much larger than the timescale of thermalization, $\tau_{\rm eq}$, over which the system of interest equilibrates with its thermal environment. These processes can then be sped up by increasing the system-environment coupling, which naturally reduces $\tau_{\rm eq}$. However,  modifying the interaction also induces additional dissipation, which prohibits the non-physical possibility of performing an isothermal process arbitrarily quickly~\cite{Newman2017Performance,Perarnau2018,wiedmann2019out,newman2019quantum} (note that increasing the coupling can lead to power output enhancements~\cite{Newman2017Performance,Perarnau2018,Abiuso2019}). Given this non-trivial trade-off, the goal of this article is to develop quantum-thermodynamic protocols that smoothly modify the system-bath interaction in order to speed up an isothermal process while keeping the overall dissipation constant. This enables us to increase the power of finite-time  heat engines and refrigerators without compromising their efficiency, a well-known  challenge in thermodynamics~\cite{Schmiedl2007,Schmiedl2007b,Esposito2010,Sivak2012,Karen2013,Holubec_2016,Ma2018,CavinaSlow,abiuso2019optimal}. 

The idea of speeding up different  thermodynamic processes  by external control has received a lot of attention in the last years.  Particularly relevant are shortcuts to adiabaticity (STA), which speed up unitary (and hence closed-system) evolutions~\cite{Torrontegui2013}, making them suited to improve the adiabatic part of thermodynamic cycles~\cite{Deng2013,Campo2014,Beau2016}. 
For open quantum systems, speed-ups of the evolution to a particular target state \cite{Jing2013,Song2016,Jing2016,Vacanti2014}, such as an equilibration or thermalization process \cite{Mukherjee2013,Suri2018,dann2018shortcut,dupays2019shortcuts,alipour2019shortcuts,dann2019fast}, have also been developed.
For classical systems, such equilibration speed-ups (the so-called Engineered Swift Equilibration  \cite{Martnez2016,LeCunuder2016,Chupeau2018}) have been experimentally tested~\cite{Martnez2016,LeCunuder2016}. Furthermore, these ideas have been extended to full isothermal classical processes, so that the state remains in the desired Gibbs distribution along the whole process \cite{Vaikuntanathan2008,Li2017,Patra2017}. These ideas have also been recently applied to the optimisation of a finite-time Carnot cycle \cite{dann2019quantum},  Otto engines \cite{villazon2019swift,tobalina2019vanishing}, and refrigerators \cite{funo2019speeding}.  
In general, such speed-ups are possible by adding a time-dependent term to the Hamiltonian which, in the presence of a thermal bath, leads to a new source of dissipation. Indeed, speed-ups of equilibration and thermalization generally come with an extra work cost \cite{Martnez2016,dann2018shortcut,tobalina2019vanishing,dann2019fast} (see also the discussion in \cite{boyd2018shortcuts} for thermodynamic computing). 

Here, our aim is to design speed-ups to isothermal processes which do not come at the price of  higher dissipation or work cost. 
 As a consequence, our speed-ups to isothermality (SI) can be readily used to maximize the power of finite-time Carnot  engines \cite{Schmiedl2007,Esposito2010,CavinaSlow,abiuso2019optimal,Ma2018} and refrigerators \cite{Wang2012,Hu2013,Hernandez_2015,holubec2020maximum,Brunner2014,Maslennikov2019,Clivaz2019} while keeping their efficiency constant. 
Due to the extra control of the system-bath interaction we find that the dissipation $W_{\rm diss}$ for  optimal SI can asymptotically decay as 
\begin{align}
W_{\rm diss} \propto \frac{1}{\tau_{\rm tot}^{2\alpha+1}}
\label{eq:decaySI1}
\end{align}
where $\tau_{\rm tot}$ is the total time of the process,  and different $\alpha > 0$ can be obtained by a suitable SI. In particular, we provide two explicit examples where $\alpha = 1/2$. The decay in Eq. \eqref{eq:decaySI1} can substantially outperform the standard scaling for large $\tau_{\rm tot}$,  $W_{\rm diss} \propto \tau^{-1}_{\rm tot}$, commonly found in  protocols where no control on the system-bath interaction is possible~\cite{Schmiedl2007b,Sivak2012,Schmiedl2007,Esposito2010,Mandal2016,Cavina2017,Abiuso2019,ma2019experimental}. Furthermore, we  show how the scaling in Eq.~\eqref{eq:decaySI1} leads to a new family of efficiencies at maximum power that interpolate between the Curzon-Ahlborn efficiency (for $\alpha=0$) and the Carnot efficiency (for $\alpha \rightarrow \infty$), see also Ref. \cite{Yang2013}. Analogous enhancements are obtained for the coefficient of performance of refrigerators \cite{Wang2012,Hu2013,Hernandez_2015}.

These results are obtained through a two-fold approach. First, we  analytically derive protocols for speeding up isothermal processes 
by assuming both  slow driving  (i.e. the timescale of the driving is slower than the time-dependent equilibration timescale) and  that the (time-dependent) coupling $g$ remains weak but non-negligible along the whole process. Second, the above approximate but analytical  approach is supported by explicit calculations for two general models of dissipation, covering both bosonic and fermionic baths. In particular, we consider quantum Brownian motion~\cite{Caldeira_Legget,Caldeira1983}, where a quantum harmonic oscillator with time-dependent frequency interacts with a time-dependent coupling to a large (but finite) set of bosonic modes, and  the resonant-level model~\cite{Ludovico2014, Esposito2015a, Esposito2015, Bruch2016, Haughian2018} where a single fermionic level with a time-dependent energy couples to an infinite bath of fermionic modes via a time-dependent interaction. By employing exact non-perturbative approaches to simulate such systems, we explicitly evaluate all sources of dissipation, including those introduced by the time-dependent system-bath interactions away from weak coupling. These complementary analyses confirm our analytical findings based on heuristic assumptions, and show that such ideas can be applied beyond the regime of weak coupling and slow driving.

As an application of our results, we demonstrate that the time of a Carnot-like engine or refrigerator cycle can be significantly reduced without increasing the dissipation by controlling the system-bath coupling appropriately, such that both power and efficiency can be \textit{simultaneously} improved. Our protocol could thus enhance the performance of quantum thermal machines in systems where the system-reservoir coupling can be controlled. We identify and discuss two promising experimental platforms, namely impurities immersed in a ultracold gases~\cite{Catani2012,Cetina2015,Hohmann2016} and mesoscopic electronic devices~\cite{Koski2014a,Koski2014b,Koski2015,Josefsson_2018}. However, numerous other possibilities can be envisaged which leverage reservoir engineering techniques, such as trapped ions~\cite{Rossnagel2016,Lemmer2018,Lindenfels2019}, superconducting circuits~\cite{Leppakangas2018} and nanomechanical systems~\cite{Klaers2017,Abdi2018}. We also show that our SI protocols are robust against control errors.

The paper is structured as follows. In Sec.~\ref{sec:quasi_static_process}, we introduce the basic  tools needed to describe isothermal processes. In Sec.~\ref{sec:speedups}, we develop the speed-ups to isothermality (SI) and optimise them to find the scaling in Eq.~\eqref{eq:decaySI1}. In Sec.~\ref{sec:EMP}, we use our findings from the previous sections to derive the efficiency at maximum power for general decays as in Eq.~\eqref{eq:decaySI1}. In Sec.~\ref{sec:numerical_results} we illustrate these general considerations numerically for quantum Brownian motion and the resonant-level model. In Sec.~\ref{sec:feasibility} we demonstrate the experimental feasibility and robustness of our proposal. We finally conclude in Sec.~\ref{sec:conclusions}.

\section{Isothermal processes} \label{sec:quasi_static_process}

Consider a driven Hamiltonian 
\begin{align}
H(t) = H^{(S)} (t) + g(t) V + \bath{H}
\label{eq:Ht}
\end{align}
where $H^{(S)}(t)$ is the Hamiltonian of the system S, on which one has experimental control, while $H^{(B)}$ is the Hamiltonian of the bath B and $V$ is the (possibly time-dependent) interaction between the two, whose strength is governed by the parameter $g$.  
The whole information of system and bath together (SB) is contained in the density matrix $\rho$.

Consider a  transformation between an initial Hamiltonian $H(0)= H^{i}$ and final one $H(\tau_{\rm tot})= H^{f}$. Without loss of generality we can normalise the parameter $t$ to the unit interval by introducing the compact notation $X_s \equiv X(s \tau_{\rm tot})$ with $s \in [0,1]$, $\tau_{\rm tot}$ the duration of the process under consideration and $X=H$, $H^{(S)}$, $\rho$ etc. The average work associated to this transformation is given by the expression
\begin{align}\label{work_def}
W=\int_0 ^1 \text{d}s\, \Tr\left(\rho_{s} \dot{H}_{s} \right) ,
\end{align}
where $\rho_{s}$ describes the instantaneous state of SB. 

Suppose first that the integrand is well described by the equilibrium value at all times, i.e. $\Tr\left(\rho_{s} \dot{H}_{s} \right)=\Tr\left(\rho_s^{\rm th} \dot{H}_{s} \right)$  with
\begin{align}
\label{eq:thermalstate}
\rho_s^{\rm th} \equiv \frac{e^{-\beta H_s}}{\mathcal{Z}},
\end{align}
and $\mathcal{Z}=\Tr(e^{-\beta H_s})$. It follows that
\begin{align}
W=\int_0^1 \text{d}s\, \Tr\left(\rho_s^{\rm th} \dot{H}_{s} \right)  =  \frac{1}{\beta} \ln \frac{\mathcal{Z}_i}{ \mathcal{Z}_f} =: \Delta F,
\label{eq:Wiso}
\end{align}
where $\mathcal{Z}$ is the partition function, $\mathcal{Z}_{i/f}=\Tr(e^{-\beta H^{i/f}})$, and $F = - \frac{1}{\beta} \log (\mathcal{Z})$ is the free energy of SB.
Eq.~\eqref{eq:Wiso} is fulfilled in the limit $\tau_{\rm tot} \rightarrow\infty$ and when the driven observables $\dot{H}$ thermalize (as expected for local observables).
Note that the quantities in Eq.~\eqref{eq:Wiso} depend on the Hamiltonian~\eqref{eq:Ht} of the system and bath together, in general. 

In the slow driving limit, i.e. for large but finite $\tau_{\rm tot}$, Eq.~\eqref{eq:Wiso} no longer holds as some work is dissipated into the bath because $\rho_s \neq \rho_s^{\rm th} $ along the trajectory.
In order to quantify the dissipated work, one introduces $W_{\rm diss}\equiv W-\Delta F \geq 0$, which tends to zero as $\tau_{\rm tot} \rightarrow \infty$.
Expanding $W_{\rm diss}$  in powers of $1/\tau_{\rm tot}$, one obtains at first order in $1/\tau_{\rm tot}$ (this corresponds to the  linear-response regime with respect to the driving speed)
\begin{align}
\label{eq:WdissIII}
W_{\rm diss}=\frac{1}{\tau_{\rm tot}}\, \int_{0}^{1}\text{d}s\,G_{\rho_s^{\rm th}}\hspace{-1mm}\left({\dot H_{t} , \dot{H}_{s}}\right)+\mathcal{O}\left(\frac{1}{\tau_{\rm tot}^2} \right)
\end{align}
 where $G_{\rho_s^{\rm th}}$ is a bilinear form evaluated at equilibrium~$\rho_s^{\rm th}$. 
 The form $G_{\rho_s^{\rm th}}$ was previously studied in different contexts. It was obtained through  linear-response theory~\cite{Sivak2012,Campisi2012geometric,acconcia2015shortcuts,Ludovico2016adiabatic}, by master equation approaches~\cite{Zulkowski2015,Cavina2017,Scandi,miller2019work}, or directly from the partition function~\cite{Nulton1,Crooks}.  For clarity of the exposition, here we focus on the latter, but our (heuristic) arguments can be extended to more general $G_{\rho_s^{\rm th}}$ (see Appendix~\ref{sec:extensions}). Furthermore, for this work it is enough to consider time-dependent Hamiltonians satisfying  $H_s = \dot{\lambda}_s \tilde{H}$, where $\tilde{H}$ is some (time-independent) observable and $\lambda_s$ is the control parameter. In this case, we can write~\cite{Nulton1,Crooks,Scandi}
 \begin{align} \label{eq:Wdiss}
W_{\rm diss}=\frac{\tau_{\rm eq}\beta}{\tau_{\rm tot} }\, \int_{0}^{1}\text{d}s\,\dot{\lambda}_s^2 {\rm cov}_{\rho_s^{\rm th}} \left(  \tilde{H} , \tilde{H} \right)+\mathcal{O}\left(\frac{\tau^2_{\rm eq}}{\tau_{\rm tot}^2} \right),
\end{align}
where $\tau_{\rm eq}$ is the timescale of relaxation (associated to $\Tr(\rho_s \tilde{H})$), and 
\begin{align}
\label{thermolength}
{\rm cov}_{\rho_s^{\rm th}}\left(  \tilde{H} , \tilde{H} \right)=\frac{1}{\beta^2} \frac{\partial^2 \ln \mathcal{Z}}{\partial \lambda^2},
\end{align}
 which can be expressed in terms of the generalised covariance defined as 
\begin{align} \label{eq:def_covariance}
&{\rm cov}_{\rho_s^{\rm th}} \left(  A , B\right) \notag \\ & =\Tr\left(A  \int_0^1 \hspace{-0.5mm}{\rm d}y \hspace{1mm}  (\rho_s^{\rm th})^{1-y} \left(B - \Tr(\rho_s^{\rm th} B)\id\right) (\rho_s^{\rm th})^{y}  \right).
\end{align}
Eq.~\eqref{thermolength} (and hence Eq.~\eqref{eq:Wdiss}) gives the standard notion of a thermodynamic metric commonly used to describe dissipative systems near equilibrium \cite{Nulton1,Crooks}, and    it provides us with a simple analytical form which depends on a single timescale $\tau_{\rm eq}$.

In this work we are interested in modifications of the system-bath interaction strength $g$, assuming initially weak coupling. In this regime, we can expand  around $g=0$, corresponding to replacing the thermal state of the interacting system $\rho_s^{\rm th}$ by the non interacting one $\rho_0^{\rm th}$.
In particular, for ${\rm cov}_{\rho_s^{\rm th}} \left(  A , A\right)$ we have
\begin{align}
{\rm cov}_{\rho_s^{\rm th}} \left(  A , A\right)= c^{(0)}_A+c^{(1)}_Ag+c^{(2)}_A g^2+..., 
\label{expansionCov}
\end{align}
where we note that a similar expansion can be performed for the more general $G_{\rho_s^{\rm th}}$ in Eq.~\eqref{eq:WdissIII}. 
 We also assume that  the thermalization time $\tau_{\rm eq} (g)$ is related to the strength of the interaction $g$ introduced in Eq.~\eqref{eq:Ht} via 
\begin{align}
\tau_{\rm eq} (g) \propto \frac{1}{g^2},
\label{eq:scalingtau}
\end{align}
which is expected in common dissipative evolutions~\cite{Breuer_Petruccione}.

Given Eqs.~\eqref{eq:Wdiss}  and \eqref{eq:scalingtau}, it is clear that the dissipated work $W_{\rm diss}$ may be reduced by increasing $g$ and hence decreasing  the thermalization timescale $\tau_{\rm eq} (g)$. However, any modification of the Hamiltonian will require additional work to be performed, leading to a non-trivial trade-off between speed and dissipated work. In what follows we develop strategies to optimally modulate $g_s$ in order to speed up the process while keeping the overall dissipation constant.

\section{Speed-ups to isothermality}\label{sec:speedups}

\begin{figure} 
	\hspace*{-0.75cm}	
	\includegraphics[scale=.6]{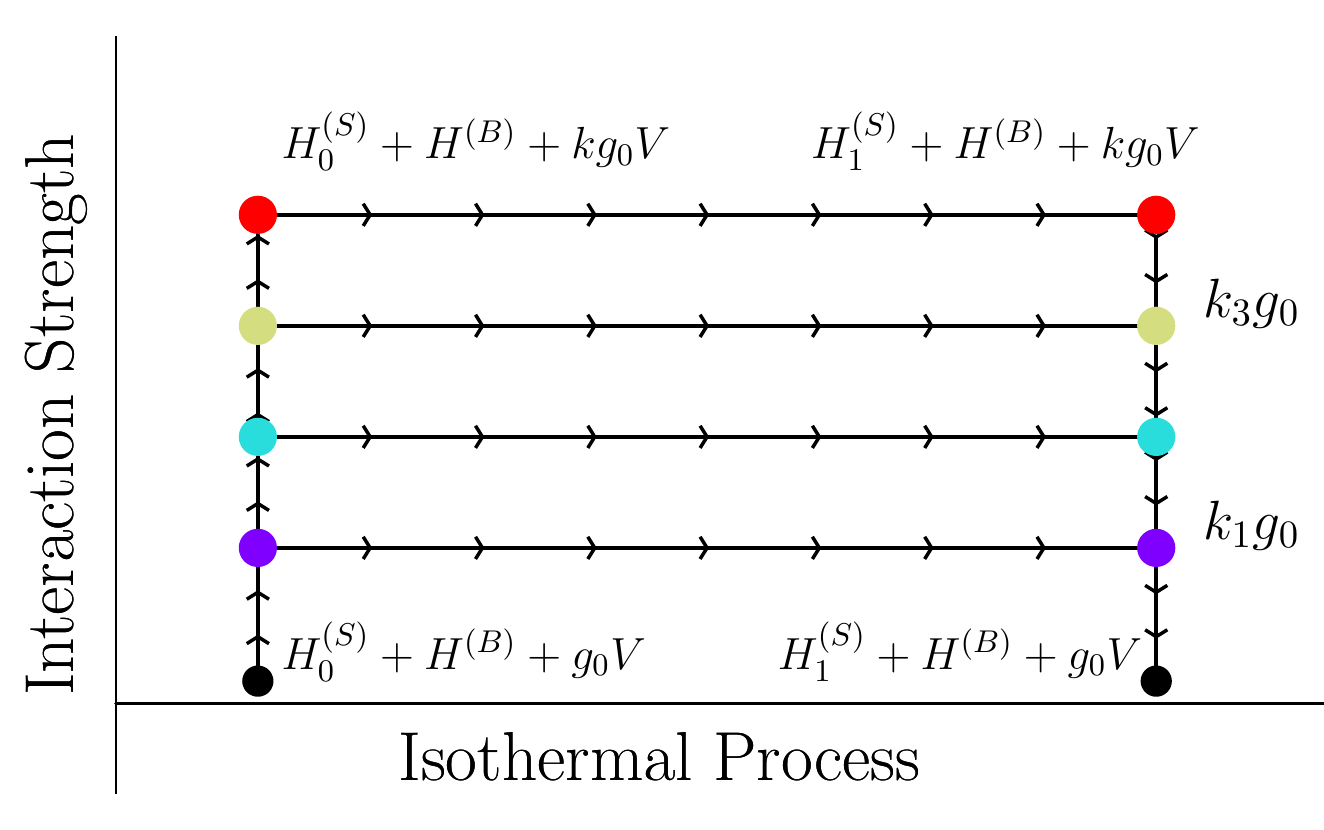}
	\caption{{\bf Schematic representation of the protocol.} We consider the family of thermodynamic protocols from an initial Hamiltonian $H_0 = H^{(S)}_0 + \bath{H} + g_0 V$ to a final Hamiltonian $H_1 = H^{(S)}_1 + \bath{H} + g_0 V$. For each protocol we choose an interaction strength $g_f = k_i g_0$.}
	\label{fig:protocol_diagram}
\end{figure}

Let us rewrite \eqref{eq:Ht} in terms of the adimensional parameter $s\in [0,1]$:
\begin{align}
\label{eq:Htp}
H_s = H^{(S)}_s + g_s V + H^{(B)},
\end{align}
where both $H^{(S)}_t$ and the interaction strength $g_s$ can be externally controlled. We focus on protocols comprising the following three steps:
\begin{enumerate}
\item[1.]  The interaction between system and bath is increased from $g_0$ to $g_f$ in a time $\tau_{\rm on}$, keeping the system Hamiltonian constant.
\item[2.] An isothermal transformation $H^{(S)}_0 \rightarrow H^{(S)}_1$ is performed in a time~$\tau_{\rm iso}$, while the interaction strength is kept constant at $g_f$.
\item[3.]  The interaction between the system and the bath is reduced to the initial value $g_0$ in a time $\tau_{\rm off}$, again holding $H^{(S)}$ constant.
\end{enumerate}
In Fig.~\ref{fig:protocol_diagram} we give a schematic representation of the thermodynamic protocol for different coupling strengths.

For simplicity we assume $\tau_{\rm off}=\tau_{\rm on}$. Since both $\tau_{\rm on}$ and $\tau_{\rm iso}$ are finite, work is dissipated during each step of the protocol. We call $\Wdis^{\rm on}$, $\Wdis^{\rm iso}$ and $\Wdis^{\rm off}$ the dissipated work in steps 1, 2 and 3, respectively. The total dissipation reads  $W_{\rm diss}=\Wdis^{\rm on}+\Wdis^{\rm iso}+\Wdis^{\rm off}$
and the total duration of the protocol is given by
\begin{align}
\label{eq:totaltime}
\tau_{\rm tot}=\tau_{\rm on}+\tau_{\rm iso}+\tau_{\rm off}=2\tau_{\rm on}+\tau_{\rm iso}.
\end{align}  
Our goal is to optimize $\tau_{\rm on}$, $\tau_{\rm iso}$ and the interaction strength $g_f$ such that $\tau_{\rm tot}$  is reduced and the dissipated work $W_{\rm diss}$ stays (approximately) constant.

\subsection{Steps 1 and 3 : Taming the dissipation when the interaction is increased or decreased} \label{sec:taming_dissipation}

We consider a family of protocols where the system-bath interaction strength changes polynomially in time according to
\begin{align}
g_s = g_i + (g_f - g_i) s^\alpha,
\label{eq:gpol}
\end{align}
for $\alpha>1$ and $s\in [0,1]$. Throughout this section, we take $g_i = g_0$ and $g_f = kg_0$, with $g_0 > 0$ a reference (weak) coupling strength. The assumption of a small, non-zero initial interaction strength is technically necessary here to ensure that we remain in the slow-driving regime, i.e. $\tau_{\rm eq} (g_s)/\tau_{\rm on} \ll 1$ $\forall s$.  However, we will later see in numerical simulations that taking $g_i=0$ leads to similar results. 

In order to quantify the dissipation during the transformation we make use of an expansion analogous to Eq.~\eqref{eq:Wdiss}, which is valid in the linear-response regime,
\begin{align} \label{eq:boundWdis_tm}
W_{\rm diss}^{\rm on} &= \frac{\beta}{\tau_{\rm on}}\int_0 ^1 \text{d}s\,\tau_{\rm eq}(g_s)  \hspace{1mm} \dot{g}_s^2 {\rm cov}_{\rho_s^{\rm th}}(V, V),
\end{align} 
where we have introduced a time-dependent equilibration timescale $\tau_{\rm eq}(g_s)$. Furthermore, through Eq.~\eqref{eq:scalingtau} we have
\begin{align}\label{eq:scalingThermalisation}
\tau_{\rm eq}(g_s) = \frac{\tau_{\rm eq}(g_0)}{(1+(k-1)s^{\alpha})^2}.
\end{align}
In order to  evaluate Eq.~\eqref{eq:boundWdis_tm}, we follow a two-fold approach. First, we approximate the covariance as in  \eqref{expansionCov} to deal with corrections of the weak coupling regime (i.e. $kg_0$ non-negligible but satisfying $kg_0\ll 1$). Second, we obtain an upper bound on  $W_{\rm diss}^{\rm on}$ using a non-perturbative approach based on the norm of $V$.  

For the first approach, we assume that $c_V^{(1)}=0$ in \eqref{expansionCov}, since $\Tr(V\rho_0^{\rm th})=0$ holds exactly in a broad class of relevant open quantum systems, such as the examples discussed in Sec.~\ref{sec:numerical_results}. Let us now consider two cases separately: keeping only the lowest-order term ($c_V^{(0)}$) or retaining also the second-order one ($c_V^{(2)}$).

\subsubsection{Zeroth order}\label{sec:zero_order}
In this case, by replacing ${\rm cov}_{\rho_s^{\rm th}}(V, V)$ by $c_V^{(0)}$ in  \eqref{eq:boundWdis_tm} we obtain
\begin{align}
W_{\rm diss}^{(1)}= \frac{\beta g_0^2 \tau_{\rm eq}(g_0)  \hspace{1mm}c_V^{(0)}}{\tau_{\rm on}} F^{(1)}(\alpha,k)
\label{disOn}
\end{align}
with
 \begin{align}
 \label{defF}
F^{(1)}(\alpha,k)=\int_0 ^1 \text{d}s\,\frac{\alpha^2 {(k-1)^2} s^{2(\alpha-1)}}{\left(1 + {(k-1)} s^{\alpha}\right)^2},
\end{align}
an integral that admits a solution in terms of the incomplete beta function. For large  $k$ (while keeping $kg_0\ll 1$ for consistency with \eqref{expansionCov}), we can approximate $F^{(1)}(\alpha,k)$ as
\begin{align}
\label{eq:largek}
F^{(1)}(\alpha,k) \approx \frac{\pi  (\alpha -1) }{\sin \left(\pi /\alpha \right)} k^{\frac{1}{\alpha }} 
\end{align}
with $\alpha >1$, whereas $F^{(1)}(1,k)= (k-1)^2/k$. This approximation, which  works  reasonably well even for low $k$,  provides an intuition of how  $F^{(1)}(\alpha,k)$ grows with $k$.

Examining Eq.~\eqref{disOn}, we see that by choosing $\tau_{\rm on}$ to be a function of $k$, such that $\tau_{\rm on}\propto F^{(1)}(\alpha,k)$, the dissipated work becomes independent of $k$. To make this more precise, we introduce $\tau_{\rm on}^{\rm weak}$ as a reference timescale for turning on the interaction to a relatively weak value with $k>1$, such that only a small amount of dissipation is incurred. For larger values of $k$, the dissipation remains small so long as the interaction is switched on over a time
\begin{align}\label{eq:polynomialChoiceTk}
\tau_{\rm on} = F^{(1)}(\alpha,k)  \tau_{\rm on}^{\rm weak}.
\end{align}
This indicates how to scale up $\tau_{\rm on}$ with $k$ in such a way that the dissipation stays constant at leading order in $g$ when the interaction is increased.
 
\subsubsection{Second order}\label{sec:second_order}
One can also consider a more conservative choice than the one in Eq.~\eqref{eq:polynomialChoiceTk} by accounting for $c_V^{(2)}$ in Eq.~\eqref{expansionCov}. The dissipation  $W_{\rm diss}^{(2)}$ induced by the second order term corresponds to
 \begin{align}
 W_{\rm diss}^{(2)}= \frac{\beta g_0^4 \tau_{\rm eq}(g_0) c_V^{(2)}}{\tau_{\rm on}} F^{(2)}(\alpha,k)
 \end{align}
with $F^{(2)}(\alpha,k)=(k-1)^2\alpha^2 / (2\alpha-1)$. For large $k$ (with $kg_0 \ll 1$), we can assume for simplicity,
\begin{align}
\label{eq:largek2}
F^{(2)}(\alpha,k) \approx  \frac{\alpha^2}{2\alpha-1} k^2.
\end{align}
Thus, by taking 
\begin{align}
\tau_{\rm on}=F^{(2)}(\alpha,k)  \tau_{\rm on}^{\rm weak},
\label{tonII}
\end{align} 
we ensure that $ W_{\rm diss}^{(2)}$ is independent of $k$. Note that this choice is more conservative since $F^{(2)}(\alpha,k) \geq F^{(1)}(\alpha,k)$ for $k\geq 1$. 

\subsubsection{Beyond weak coupling: a non-perturbative   approach} \label{sec:higher_order}

In principle, one can extend the previous considerations to find  more conservative choices of $\tau_{\rm on}$ as a function of $k$ by accounting for higher orders in Eq.~\eqref{expansionCov}. However, for stronger couplings a more useful approach is to use the fact that ${\rm cov}_{\rho_s^{\rm th}} \left(  V , V\right) \leq 2||V||^2$, in order to bound the (exact) dissipation \eqref{eq:boundWdis_tm} as
\begin{align}
W_{\rm diss} \leq  \frac{\beta g_0^2 \tau_{\rm eq}(g_0)  \hspace{1mm}2||V||^2}{\tau_{\rm on}} F^{(1)}(\alpha,k).
\label{disOnIII}
\end{align}
Hence, in models where $g_0^2||V||^2$ is finite (and possibly small), it appears plausible that the choice~\eqref{eq:polynomialChoiceTk} is in fact already sufficient to keep the dissipation controlled (note that with~\eqref{eq:polynomialChoiceTk} the upper bound \label{disOnII} becomes independent of $k$). Importantly, the bound \eqref{disOnIII} also works for strongly correlated and non-Markovian systems, suggesting that our considerations  also apply  for strongly correlated systems that thermalize \cite{Eisert2015}. This will be confirmed later through exact numerical examples at strong coupling for fermionic and bosonic baths. In particular, the bound \eqref{disOnIII} can become  tight for finite-dimensional and locally interacting systems, such as fermionic or spin models, where $\| V \|$ is of the order of the system-bath boundary  and independent of the size of the bath. In such cases, the scaling of the equilibration time might differ from Eq.~\eqref{eq:scalingtau}, but our framework can be easily adapted to account for that. 

\subsubsection{Discussion}

Summarising, in this section we showed how to scale up $\tau_{\rm on}$ with $k$ to ensure that $W_{\rm diss}$ does not increase as we increase the interaction. We followed two complementary approaches. First, taking a perturbative expansion of $W_{\rm diss}$ for weak coupling, we derived two possible choices: Eq.~\eqref{eq:polynomialChoiceTk} and Eq.~\eqref{tonII}. The former ensures stays $W_{\rm diss}$ stays constant at leading order in the expansion (zeroth order), whereas the latter ensures that $W_{\rm diss}$ does not increase with $k$ up to second order in $g$. 
Second, we showed that one can also upper bound $W_{\rm diss}$  by a $k$-independent bound by combining \eqref{disOnIII} and \eqref{eq:polynomialChoiceTk}, a bound which holds at arbitrary strong coupling (i.e. large $g$) as long as $||V||$ is finite.
In Sec.~\ref{sec:numerical_results} we will test these choices for fermionic and bosonic baths (see Figs.~\ref{fig:W_diss}), showing that these generic considerations work well in relevant  physical models even at reasonably strong coupling.

\subsection{Step 2: Isothermal part of the process}

Now we focus on the isothermal part of the protocol. The protocol consists of modifying the Hamiltonian of the system $H^{(S)}_t$ whilst keeping the coupling strength $g$ constant. We introduce  $\tau_{\rm iso}^{\rm weak}$ as the time spent to perform the isothermal part of the protocol for $k=1$, i.e., in the absence of modulations of the interaction. By assuming the scaling in Eq. \eqref{eq:scalingtau} and by using the expansion in Eq. \eqref{eq:Wdiss}, we can choose the time $\tau_{\rm iso}$ of the isothermal process for $k>1$ as
\begin{align}
\tau_{\rm iso}= \frac{\tau_{\rm iso}^{\rm weak}}{k^2}
\label{eq:choiceTiso}
\end{align}
in order to keep the dissipation constant for any value of $k$. Similar to  the previous section,  this is strictly valid at leading order in $g_0$, i.e. when keeping only the first term in \eqref{expansionCov}. This appears justified in the dissipative systems we consider in this work (see Sec.~\ref{sec:numerical_results}).

\subsection{Full protocol}

\begin{figure}
\includegraphics[width=1\linewidth]{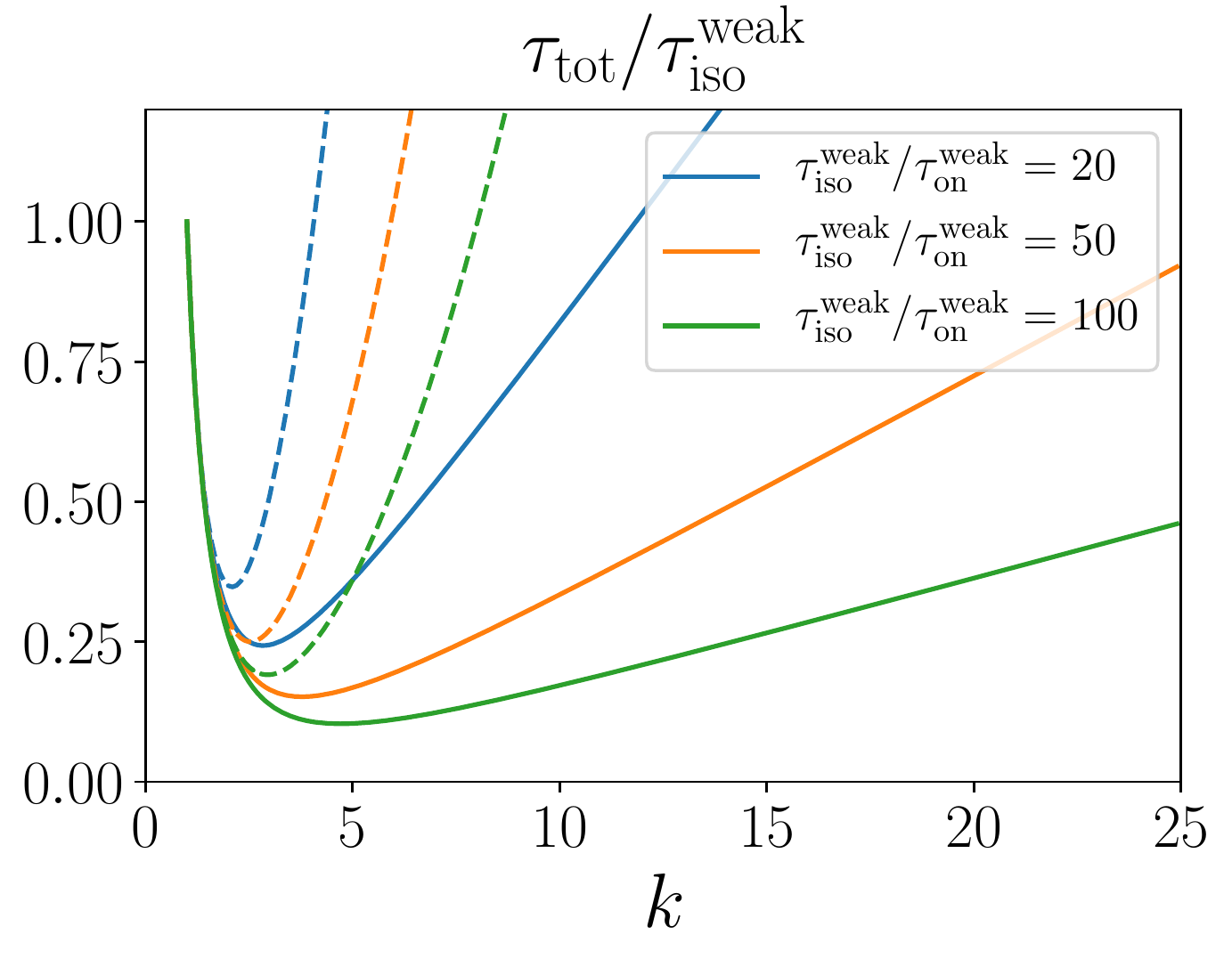}
\caption{The total time $\tau_{\rm total}$ in \eqref{eq:totalTime} as a function of $k$,  for different values of $\mathcal{T}=\tau^{\rm weak}_{\rm iso}/\tau^{\rm weak}_{\rm on}$: $\mathcal{T}=20$ (blue), $\mathcal{T}=50$ (orange), $\mathcal{T}=100$ (green); and $\alpha=1$. The dashed line corresponds to $F^{(2)}$ and the solid line to $F^{(1)}$. Results for both $F^{(2)}$ and $F^{(1)}$ are exact, the former being obtained  through the exact integral  expression \eqref{defF}. 
\label{fig:TvsKk}
}
\end{figure}

Collecting all the considerations above, we have devised choices of $\tau_{\rm on}$ and $\tau_{\rm iso}$ as a function of $k$ which guarantee an overall constant dissipation. 
The total time of the protocol reads
\begin{align}
\label{eq:totalTime}
\tau_{\rm tot}= 2F^{(i)}(\alpha,k) \tau_{\rm on}^{\rm weak} +\frac{\tau_{\rm iso}^{\rm weak}}{k^2},
\end{align}
where $F^{(i)}(\alpha,k)$ is given by either Eq.~\eqref{defF} or Eq.~\eqref{eq:largek2}, the latter being more conservative than the former in order to not increase the dissipation (see Sec.~\ref{sec:numerical_results} for examples). 
In Fig. \ref{fig:TvsKk} we show the behaviour of Eq. \eqref{eq:totalTime} for different values of the dimensionless ratio $\mathcal{T}=\tau^{\rm weak}_{\rm iso}/\tau^{\rm weak}_{\rm on}$, and for both choices  \eqref{defF} and \eqref{eq:largek2}. Note that for large $\mathcal{T}$, as in realistic situations (normally the isothermal process is much longer than the time spent switching the system-bath interaction on and off), we obtain substantial improvements on the time of protocol. Our proposal hence provides a way of substantially speeding up isothermal processes through control of the system-bath interaction, which crucially does not come at the price of increased dissipation or work cost. 

\subsection{Optimal protocols and decay of dissipation}\label{sec:optimalscaling}

Above, we designed a family of protocols in which the dissipation remains constant, while the total time of the process can be adjusted as a function of $k$ (see also Fig. \ref{fig:TvsKk}). 
Let us now minimize the expression in Eq.~\eqref{eq:totalTime} to find the fastest isothermal process for a given dissipation. 

We first consider the zeroth-order expansion from Sec.~\ref{sec:zero_order}. To obtain an analytical expression, we use the large-$k$ approximation in Eq.~\eqref{eq:largek} to obtain
\begin{align}
\tau_{\rm tot} = 2 D_\alpha k^{\frac{1}{\alpha }} \tau_{\rm on}^{\rm weak} +\frac{\tau_{\rm iso}^{\rm weak}}{k^2},
\label{Texds}
\end{align}
where $D_\alpha=\pi  (\alpha -1) \csc \left(\pi /\alpha \right)$.
This expression can be minimized with respect to $k$, yielding
\begin{align}
k= \left(\frac{\alpha   \tau^{\rm weak}_{\text{iso}}}{D_\alpha \tau^{\rm weak}_{\text{on}}}\right){}^{\frac{\alpha }{2 \alpha +1}},
\label{eq:kchoice}
\end{align}
and the corresponding minimal time
\begin{align}
\tau_{\rm tot} & = C_\alpha \tau_{\rm iso}^{\rm weak} \left(\frac{\tau_{\rm on}^{\rm weak}}{\tau_{\rm iso}^{\rm weak}}\right)^{\frac{2\alpha}{2\alpha+1}}\notag 
\label{eq:totOptTime}
\end{align}
where $C_{\alpha }$ is the constant
\begin{align}
C_{\alpha}=(2\alpha+1)\left(\frac{D_{\alpha}}{\alpha}\right)^{\frac{2\alpha}{2\alpha+1}}.
\end{align}

For a standard isothermal process at $k=1$ in the weak coupling regime, in which the interaction is not modified,  at leading order in $1/\tau^{\rm weak}_{\text{iso}}$ the dissipated work can be expressed as~\cite{Esposito2010,CavinaSlow,Scandi}  
\begin{align}
W_{\rm diss}^{ \rm weak}= \frac{\Sigma}{\tau^{\rm weak}_{\text{iso}}}
\label{eq:Wdisweakk}
\end{align}
where $\Sigma>0$ can be obtained from the integral expression in Eq.~\eqref{eq:Wdiss}, where we note we have neglected the cost of turning on/off the interaction due to the weak coupling (the importance of this assumption will be discussed in more detail in Sec.~\ref{sec:numerical_results}, where all work costs are evaluated explicitly).
By construction, the family of protocols in Eq.~\eqref{eq:totOptTime} will dissipate the same $W_{\rm diss}=W_{\rm diss}^{ \rm weak}$. 
If we combine this observation with Eq.~\eqref{eq:totOptTime}  and \eqref{eq:Wdisweakk}, we  obtain that
\begin{align} 
W_{\rm diss} = \Sigma  C_{\alpha}^{2\alpha +1} \frac{(\tau_{\rm on}^{\rm weak})^{2\alpha}}{\tau^{2\alpha +1}_{\rm tot}},
\label{eq:Wdisopt}
\end{align}
with $\alpha>0$.  For constant $\tau_{\rm on}^{\rm weak}$, the dissipation decays as $\tau_{\rm tot}^{-(2\alpha+1)}$ in the total time $\tau_{\rm tot}$ of the process, which can greatly outperform the standard decay in Eq.~\eqref{eq:Wdisweakk}.  

 Naively, the decay in Eq. \eqref{eq:Wdisopt} may suggest that one can make the dissipation arbitrary small simply by increasing $\alpha$. This is not the case, however, due to the contribution of the constant $C_{\alpha}$, which diverges exponentially as $\alpha$ increases. As a consequence 
 one can show that for any $\tau$ there exists an optimal $\alpha$, which scales logarithmically in $\tau$. Hence, one needs exponentially long protocols in order to choose larger $\alpha$. 

Next, we discuss the case where $\tau_{\rm on}$ is scaled as in Eq.~\eqref{eq:largek2}, in order to account for contributions to the dissipated work at second order in $g$. Using the large-$k$ approximation in Eq.~\eqref{eq:largek2}, the total time now reads as
\begin{equation}
\label{eq:tau_tot_ksqd}
    \tau_{\rm tot} = 2B_\alpha \tau_{\rm on}^{\rm weak} k^2 +\frac{\tau_{\rm iso}^{\rm weak}}{k^2},
\end{equation}
where $B_\alpha = \alpha^2/(2\alpha-1)$. Following the same steps as before, we find that the total time is minimized when we choose $k^4 = \tau_{\rm iso}^{\rm weak}/(2B_\alpha\tau_{\rm on}^{\rm weak})$, yielding $\tau_{\rm tot}=\sqrt{8B_\alpha \tau_{\rm iso}^{\rm weak} \tau_{\rm on}^{\rm weak}}$. This leads to a decay of the efficiency given by
\begin{align} 
W_{\rm diss} = \frac{8\alpha^2\Sigma \tau_{\rm on}^{\rm weak} }{(2\alpha-1) \tau^{2}_{\rm tot}}.
\label{eq:WdisoptIII}
\end{align}
Therefore, for a fixed $\tau_{\rm on}^{\rm weak}$, the dissipation decays with the total time as $\tau^{-2}_{\rm tot}$, in contrast to the standard decay~\eqref{eq:Wdisweakk}.

\section{Efficiency at maximum power through speed-ups to isothermality}\label{sec:EMP}

In this section, we study the implications of optimal shortcuts to isothermality for thermodynamic cycles. In Sec.~\ref{sec:speedups}, we presented different possible protocols for speeding up an isothermal process.  Here, we  use a general form for the decay of the dissipated work which encompasses all regimes considered in Sec.~\ref{sec:speedups}. Indeed, let us assume that the dissipation decays as 
\begin{align}
W_{\rm diss} = \frac{\Sigma_\gamma}{\tau_{\rm tot}^\gamma}
\label{eq:WdisoptII}
\end{align}
where $\gamma\geq 1$ and $\tau_{\rm tot}$ is the time of the  process. For the optimal shortcuts to isothermality, we have that $\gamma=2\alpha+1$ with $\alpha>0$ and $\Sigma_\gamma=\Sigma C_{\alpha}^{2\alpha+1}(\tau_{\rm on}^{\rm weak})^{2\alpha}$, whereas for the more conservative choice in Eq.~\eqref{eq:largek2} we have $\gamma=2$ and $\Sigma_\gamma$ given in Eq.~\eqref{eq:WdisoptIII}.

\subsection{Heat engines}

 We consider a finite-time Carnot-like cycle between two thermal baths at different temperatures $T_h$ and $T_c$~\cite{Esposito2010,Hernandez_2015,CavinaSlow,abiuso2019optimal}. Furthermore, when the (finite-time) isothermal part of the cycle is carried out, we assume a decay as in Eq.~\eqref{eq:WdisoptII}. 
 Using $Q+W=\Delta E_S$,  the heat exchanged between the system and each of the two thermal baths reads
\begin{align}
Q_c & = T_c \left(-\Delta S-\frac{\Sigma_\gamma}{\tau_{c}^{\gamma}} +...\right),\nonumber\\
Q_h & =  T_h \left(\Delta S-\frac{\Sigma_\gamma}{\tau_{h}^{\gamma}} +...\right),
\label{eq:heatDefinition}
\end{align}
where $\tau_{c,h}$ are the times of the isothermal processes (with the cold, hot bath, respectively), and 
we have assumed a symmetric cycle such that the constants $\Sigma_\gamma$ are equal for each isothermal process~\cite{Esposito2010,CavinaSlow,abiuso2019optimal}.
The efficiency of the engine is given by
 \begin{align}
\eta = 1+\frac{Q_c}{Q_h},
\end{align}
 whereas the power reads
\begin{align}
P = \frac{Q_h+Q_c}{\tau_h + \tau_c}. 
\label{PowerHeatEngine}
\end{align}

\begin{figure}
{\hspace*{-.5cm}
\includegraphics[width=\linewidth]{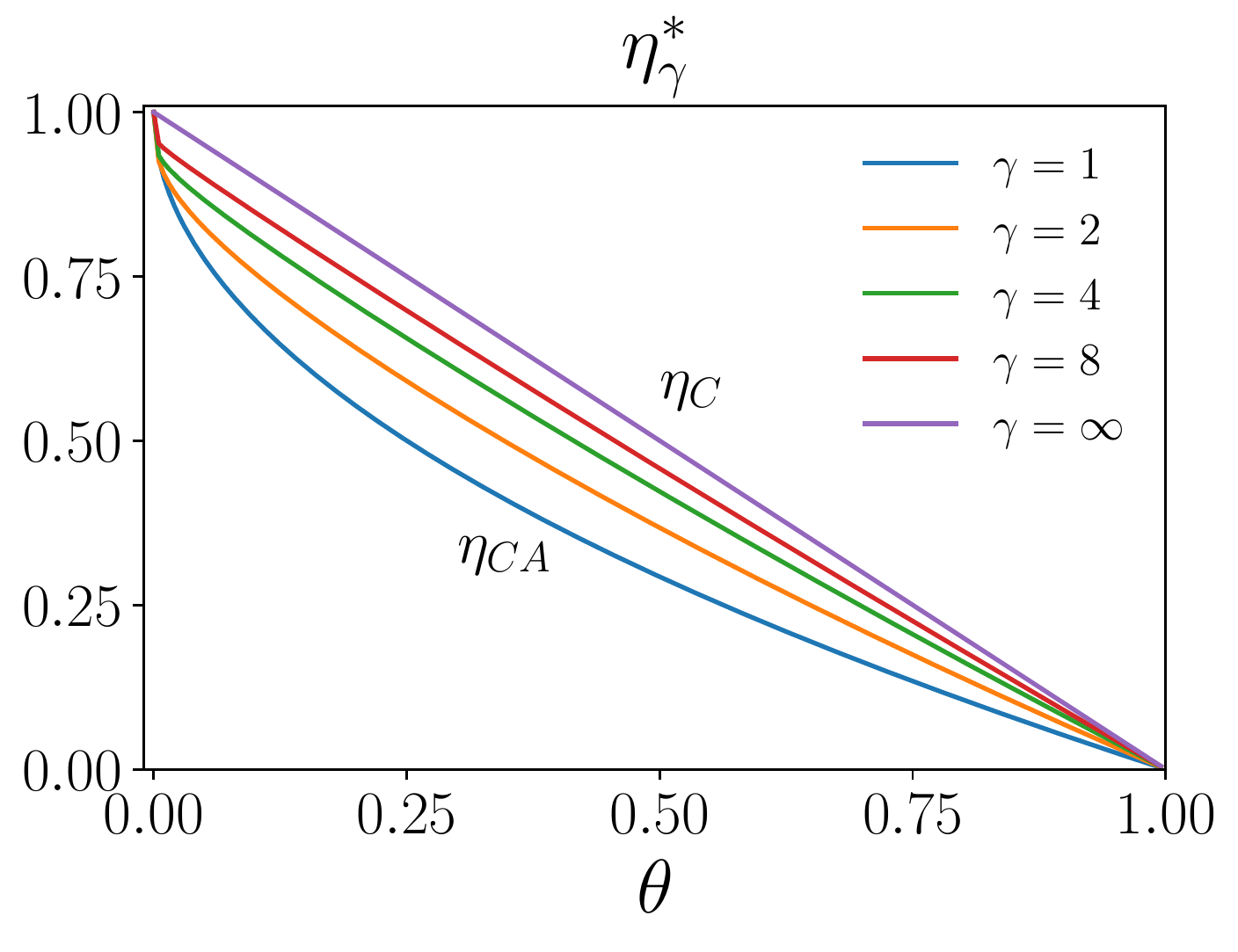}
\caption{{\bf Efficiency at maximum power.} The figure shows $\eta^*_\gamma$ for $\gamma=1,2,4,8,\infty$; with $\gamma=1$ and $\gamma =\infty$ corresponding to Curzon-Ahlborn ($\eta_{CA}$) and Carnot efficiency ($\eta_C$), respectively. }
\label{fig:EMP}
}
\end{figure}

In the case of $\gamma=1$, i.e. $W_{\rm diss}\propto \tau_{\rm tot}^{-1}$, the efficiency at maximum power $\eta^*$ is given by the Curzon-Ahlborn efficiency~\cite{Schmiedl2007,Esposito2010}. We want to compute $\eta^*$ for a generic value of $\gamma$; see also  Ref.~\cite{Yang2013} for a similar analysis. 
The maximum power is obtained by imposing the two conditions:
\begin{align}   \label{eq:minConditions}
\frac{\partial P}{\partial \tau_{c}}=0, \ \ \  \frac{\partial P}{\partial \tau_{h}}=0.
\end{align}
The system has a unique real and positive solution for $\tau_{c,h}$ given by:
\begin{align}
&\tau_c= \frac{\theta ^{\frac{1}{\gamma +1}} }{\theta ^{\frac{1}{\gamma +1}}+1} \left(\frac{\Sigma_{\gamma}(\gamma +1) \theta  \left(\theta ^{-\frac{1}{\gamma +1}}+1\right)^{\gamma +1}}{\Delta S  (1-\theta )}\right)^{\frac{1}{\gamma }}\nonumber\\
&\tau_h= \frac{\tau_c}{\theta ^{\frac{1}{\gamma +1}}},
\label{optimaltimes}
\end{align}
where we used the notation $\theta := T_{c}/T_{h}$. 
The corresponding efficiency at maximum power reads
\begin{align}
\eta^*_{\gamma}&=1-\frac{\theta ^{\frac{1}{\gamma +1}} \left((\gamma +1) \theta ^{\frac{\gamma }{\gamma +1}}+\gamma  \theta +1\right)}{(\gamma +1) \theta ^{\frac{1}{\gamma +1}}+\gamma +\theta },
\label{generalisedEMP}
\end{align}
which depends only on $\gamma$ and the ratio of temperatures $\theta$. 
This formula has two interesting limits: for $\gamma \rightarrow 1$, one obtains the Curzon-Ahlborn efficiency $\eta_{1}^{*}=1-\sqrt{\theta}\equiv \eta_{CA}$, while for $\gamma \rightarrow \infty$ we regain the Carnot efficiency $\eta_{\infty}^{*}= 1-\theta \equiv \eta_C$. The efficiency at maximum power interpolates between these two regimes as $\gamma$ varies, as illustrated in  Fig.~\ref{fig:EMP}. If we expand Eq.~\eqref{generalisedEMP} around $\theta \rightarrow 1$ (i.e. $\eta_C \rightarrow 0$), we obtain
\begin{align}
\eta^*_{\gamma}=\frac{\gamma  }{\gamma +1}\eta_C +\frac{\gamma}{2 (\gamma +1)^2}\eta^2_C+\mathcal{O}(\eta_C^3).
\label{generalisedEMPII}
\end{align}
The expansion in Eq.~\eqref{generalisedEMPII} neatly shows how $\eta^*_{\gamma}$ approaches   $ \eta_C$ as $\gamma$ increases. Notice that for the optimal SI that we defined in the previous section, the time of the process in Eq.~\eqref{optimaltimes} is proportional to $C_\gamma$, and hence tends to infinity as  $\eta^*_{\gamma}\rightarrow \eta_C$, preventing the possibility of achieving a Carnot cycle with finite power. 
 
\subsection{Refrigerators}

Now we consider a refrigerator, in which  input power is used to extract energy from the cold bath (i.e. to reverse the natural heat flow) \cite{Wang2012,Hu2013,Hernandez_2015,holubec2020maximum}.
The cooling power is given by:
 \begin{align}
P_{\rm c} = \frac{Q_c}{\tau_h + \tau_c},
\label{CoolingPower}
\end{align}
whereas the figure of merit corresponding to the efficiency is given by the coefficient of performance (COP), defined as:
 \begin{align}
{\rm COP} = \frac{Q_c}{W_{\rm in}}\leq {\rm COP_C},
\end{align}
where $W_{\rm in}=-Q_c - Q_h$ is the input work (minus the extracted work in \eqref{PowerHeatEngine}), and 
${\rm COP_C}$ is the Carnot COP, given by ${\rm COP_C} := (\theta^{-1}-1)^{-1}$.
As in the previous section, our goal is to extent previous results in the low-dissipation regime for the COP at maximum cooling power \cite{Wang2012,Hu2013,Hernandez_2015,holubec2020maximum} to the more generic decay of dissipation given in \eqref{eq:WdisoptII}.

\begin{figure}
{\hspace*{-.5cm}
\includegraphics[width=\linewidth]{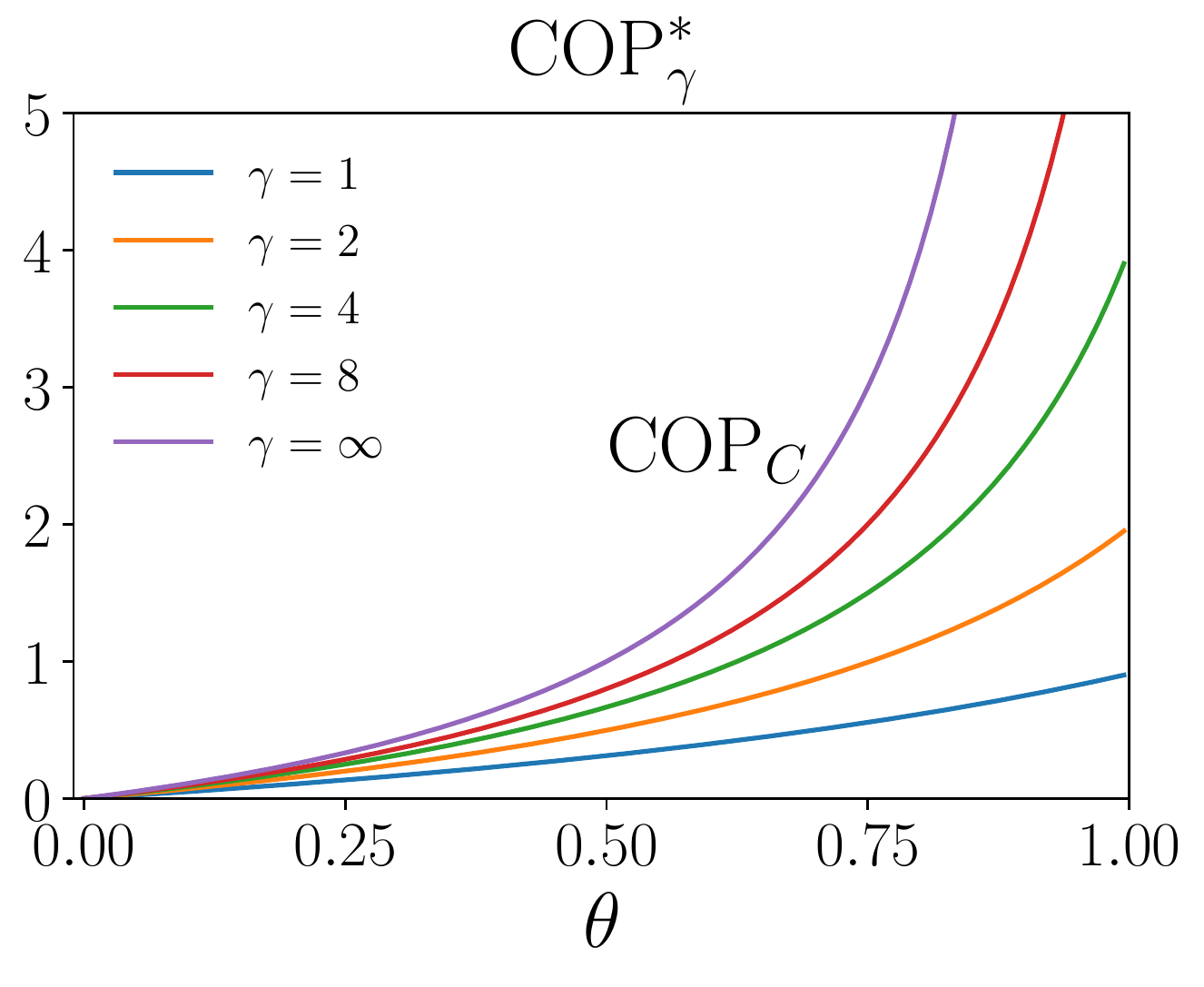}
\caption{{\bf Efficiency at maximum power.} The figure shows ${\rm COP}^*_{\gamma}$ for $\gamma=1,2,4,8,\infty$, and $R=10$; with $\gamma =\infty$ corresponding to Carnot COP (${\rm COP_C}$), respectively. }
\label{fig:REFR}
}
\end{figure}

We first note from  \eqref{CoolingPower}  that the maximum condition Eq.~\eqref{eq:minConditions} would lead to the unphysical solution $\tau_h \equiv 0$. For this reason it is convenient to maximise $P_c$ for a fixed ratio $R := \tau_h/\tau_c$.  In this case, the maximisation of \eqref{CoolingPower} also has a unique real and positive solution for $\tau_{c,h}$ given by:
\begin{align}
&\tau_c= \left(\frac{(\gamma +1) \Sigma _{\gamma }}{\text{-$\Delta $S}}\right)^{\frac{1}{\gamma }},\nonumber\\
&\tau_h= R\, \tau_c .
\label{SolutionTimeRefrig}
\end{align}
The corresponding COP at maximum power reads
\begin{align}
{\rm COP}^*_{\gamma}=\frac{1}{ \frac{1+\gamma+R^{-\gamma}}{\gamma \theta}-1},
\label{generalisedEMP}
\end{align}
which depends only on the ratio of temperatures $\theta$, $\gamma$ and $R$. 
Again, in the limit of  $\gamma \rightarrow \infty$ we regain the Carnot COP, if $R>1$, i.e.  $\tau_h> \tau_c $ (note that this condition also appears for consistency of the solution \eqref{SolutionTimeRefrig}: for $R<1$ the second term of $Q_h$ in \eqref{eq:heatDefinition} diverges hence making the power expansion in the  low-dissipation regime unjustified). The dependence of ${\rm COP}^*_{\gamma}$ on $\gamma$ is illustrated in Fig.~\ref{fig:REFR}, where it is observed how $\gamma>1$ enables higher COP at maximum power. Moreover, if we expand for $\theta\rightarrow0$ (i.e. ${\rm COP_C}\rightarrow0$), we obtain
\begin{align}
    {\rm COP}^*_{\gamma} &= \frac{1}{(1-R^{-\gamma })\gamma^{-1}+1} {\rm COP_C}+\nonumber\\&+\frac{2  \left(1-R^{\gamma }\right)}{R^{\gamma }\gamma  \left((1-R^{-\gamma })\gamma^{-1}+1\right)^2} {\rm COP_C}^2 + \bigo{{\rm COP_C}^3},
\end{align}
which also shows how $ {\rm COP}^*_{\gamma}$ continuously approaches  ${\rm COP_C}$ as $\gamma$ increases.

\section{Numerical results}\label{sec:numerical_results}

In the previous sections we have combined heuristic and rigorous arguments to show that the time of an isothermal process in Eq.~\eqref{eq:totalTime} can be considerably reduced by suitably modifying the coupling between system and bath. The goal of this section is to illustrate these considerations for exactly solvable models. Specifically, we consider two complementary examples: a bosonic environment described by the Caldeira-Leggett model and a fermionic bath described by the resonant-level model. The quadratic nature of their corresponding Hamiltonians allows us to simulate the systems exactly at arbitrarily strong coupling and driving speed, hence going beyond our previous analytical considerations.

With the Caldeira-Leggett model, we study a problem with bosonic degrees of freedom using exact calculations but with a finite, discretized bath (the bath is large enough that our statements about thermalization remain meaningful). In this context, we quantitatively demonstrate that the heuristic assumptions underlying our analytical results hold to an excellent approximation, even with relatively fast driving and coupling strength. Then we move to a resonant-level model with fermionic degrees of freedom, which is analysed using an approximate analytical approach. This allows us to study both the slow-driving and fast-driving regimes in order to demonstrate a genuine speed-up for isothermal processes at strong coupling. Finally, we consider a full Carnot cycle using the resonant fermion level as a working medium, demonstrating explicitly that SIs allow for increased power without compromising on efficiency.

\subsection{The Caldeira-Leggett model}
\label{CLmodel}

We start by illustrating our results with the Caldeira-Leggett (CL) model~\cite{Caldeira_Legget,Caldeira1983,Breuer_Petruccione}, prototypical example of a quantum Brownian motion. The CL model describes a Brownian quantum particle of mass $m$ in a harmonic potential. The full Hamiltonian consists of four terms:
\begin{align} \label{eq:Hcaldeira}
H = H^{(S)} + H^{(B)} + H^{(SB)} + H^{(R)},
\end{align}
where the Hamiltonian of the system S reads 
\begin{align} \label{eq:H_S}
H^{(S)} = \frac{1}{2} \left( m \omega_S ^2 x^2 + \frac{p^2}{m} \right) 
\end{align} 
with $x$ and $p$ the position and momentum operators. The Hamiltonian of the bath B is
\begin{align} \label{eq:CL_H_B}
H^{(B)}  = \frac{1}{2} \sum_{n = 0}^N  \left( \frac{p_n^2}{m_n}  + m_n \omega_n^2 x_n^2 \right),
 \end{align}
where $\omega_n = \frac{n}{N}(\omega_{\rm max} - \omega_{\rm min}) + \omega_{\rm min} $ are the frequencies of the modes in the bath, and we defined $\omega_{\rm max} = 2 \omega_S$ and $\omega_{\rm min} = \omega_S / N$. 
The interaction $H^{(SB)}$ between the system and the bath is defined as
 \begin{align}\label{eq:CL_H_I}
H^{(SB)} &= x \sum_n \gamma_n x_n, 
\end{align} 
where $\gamma_n$ are the coupling constant between system and bath. 
The relevant bath properties are characterised by the spectral density $\mathfrak{J(\omega)} = 2\pi \sum_n  \frac{\gamma_n^2}{\omega_n}   \delta(\omega- \omega_n)$. 
In the remainder we will assume all the masses $m, m_n = 1$, and that the couplings satisfy:  $\gamma_n = g \omega _n \sqrt{\omega_{\rm max} / (2 \pi N )}$, which leads to an Ohmic spectral density with hard cutoffs in the continuum limit $N\rightarrow \infty$ (see e.g. Appendix G of \cite{Perarnau2018}):  
\begin{equation}\label{eq:spectral_density_boson}
\mathfrak{J}(\omega) = g^2 \omega \Theta(\omega_{\rm max}-|\omega|).
\end{equation}
Here, $g$ is a time-dependent coupling strength, leading to a characteristic dissipation rate $g^2$, while $\Theta(z)$ is the Heaviside step function.
The last term $H^{(R)}$ in Eq.~\eqref{eq:Hcaldeira} is a renormalization term which ensures the positivity of~$H$,
\begin{align}
H^{(R)} = x^2 \sum_n \frac{\gamma_n^2}{m_n \omega_n^2}, 
\end{align} 
which may be absorbed within $H^{(S)}$. 

 The CL Hamiltonian in Eq. \eqref{eq:Hcaldeira} is quadratic. This enables us to diagonalize it efficiently and to describe the time-evolved state by covariance matrices (of size $2N \times 2N$ for systems composed of $N$ particles), allowing us to reach large but finite baths. Thus, the dynamics induced by the CL Hamiltonian in Eq.~\eqref{eq:Hcaldeira} can be simulated without making any assumption on the coupling strength $g$ (see e.g. \cite{Rivas2010} for details).

\begin{figure}
	\hspace*{-0.8cm}
	\subfloat{
		\begin{picture}(0,0)
		\put(50,82){\includegraphics[scale=0.32]{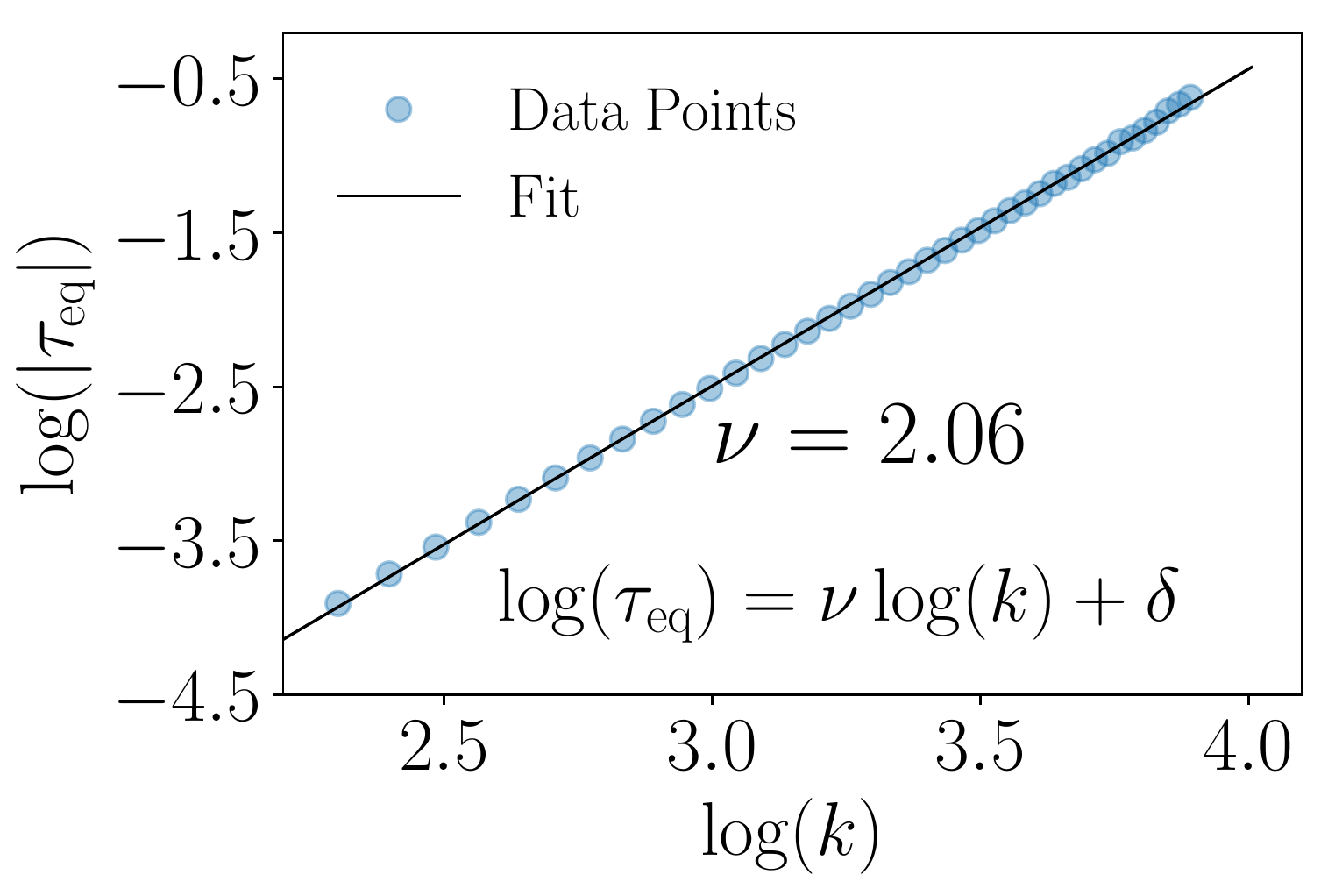}}
		\end{picture} 
	}	
	\includegraphics[width=0.5\textwidth]{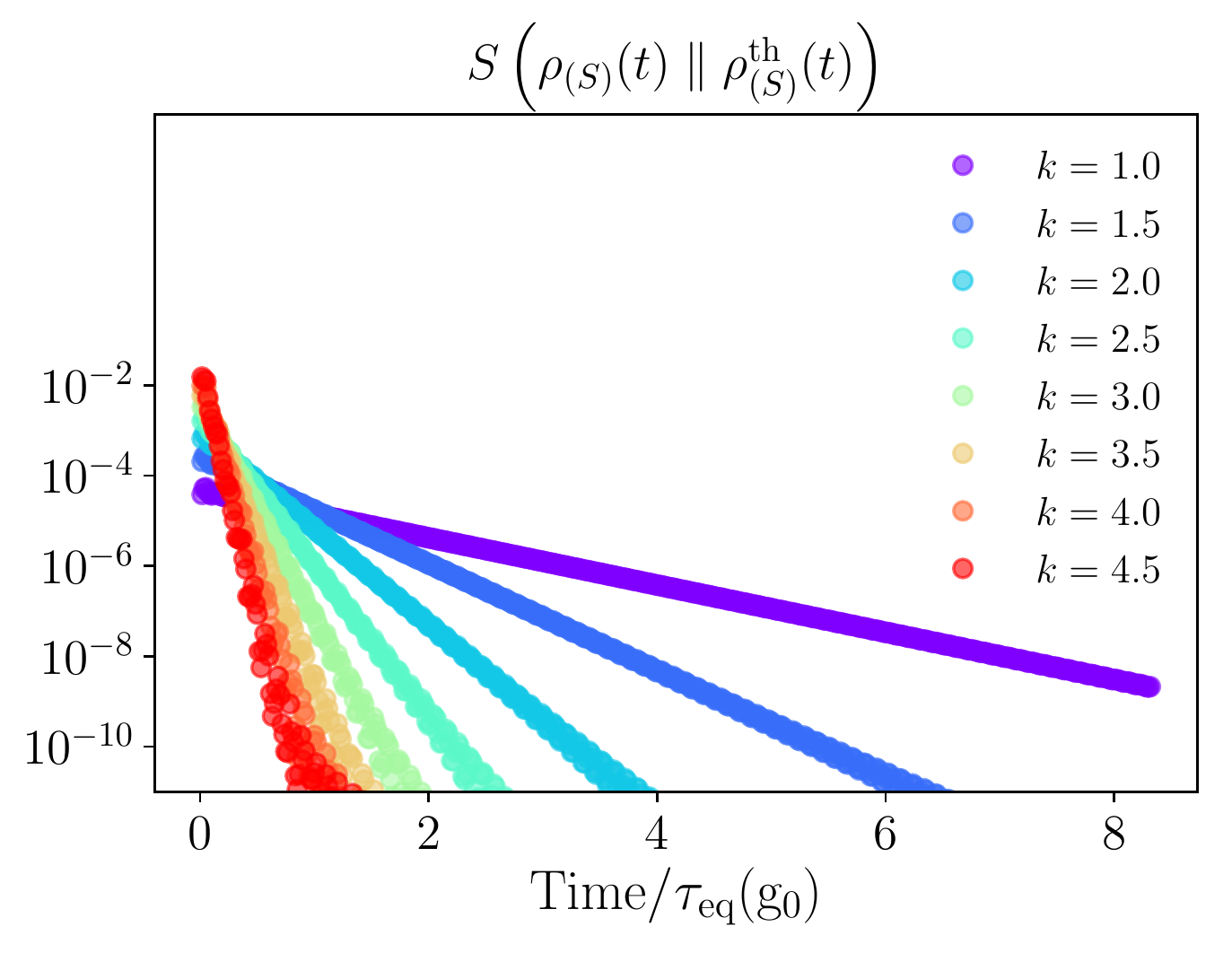}
	\caption{{\bf Relaxation to the thermal state.} We compute the relative entropy $S(\sys{\rho}(t) \parallel \rho^{\rm th} _{(S)} (t))$ between the reduced state $\rho_{(S)} (t) = \Tr_B (\rho (t))$ of the total time-dependent density matrix and thermal state of the system $\rho^{\rm th} _{(S)} (t)$. For a wide range of values of $k$ the time evolved state approaches the thermal equilibrium exponentially $S(\sys{\rho}(t) \parallel \rho^{\rm th} _{(S)} (t)) \sim e^{-t / \tau_{\rm eq} (g)}$. {\bf Inset:} We extrapolate the decay of the relaxation time with a power law $\tau_{\rm eq} (g) \sim \tau_{\rm eq} (g_{0}) {k}^{-\nu}$. The optimal fit corresponds to $\nu = 2.06$ and $g_0^2 \tau_{\rm eq} (g_{0}) = 0.59$. Parameters: $N=300$ and $g_0^2/\omega_S =0.01$.
	}
	\label{fig:Hs_therm}
\end{figure}

\begin{figure} 
	\hspace*{-.8cm}
	\subfloat{
		\begin{picture}(0,0)
		\put(45,82){\includegraphics[scale=0.32]{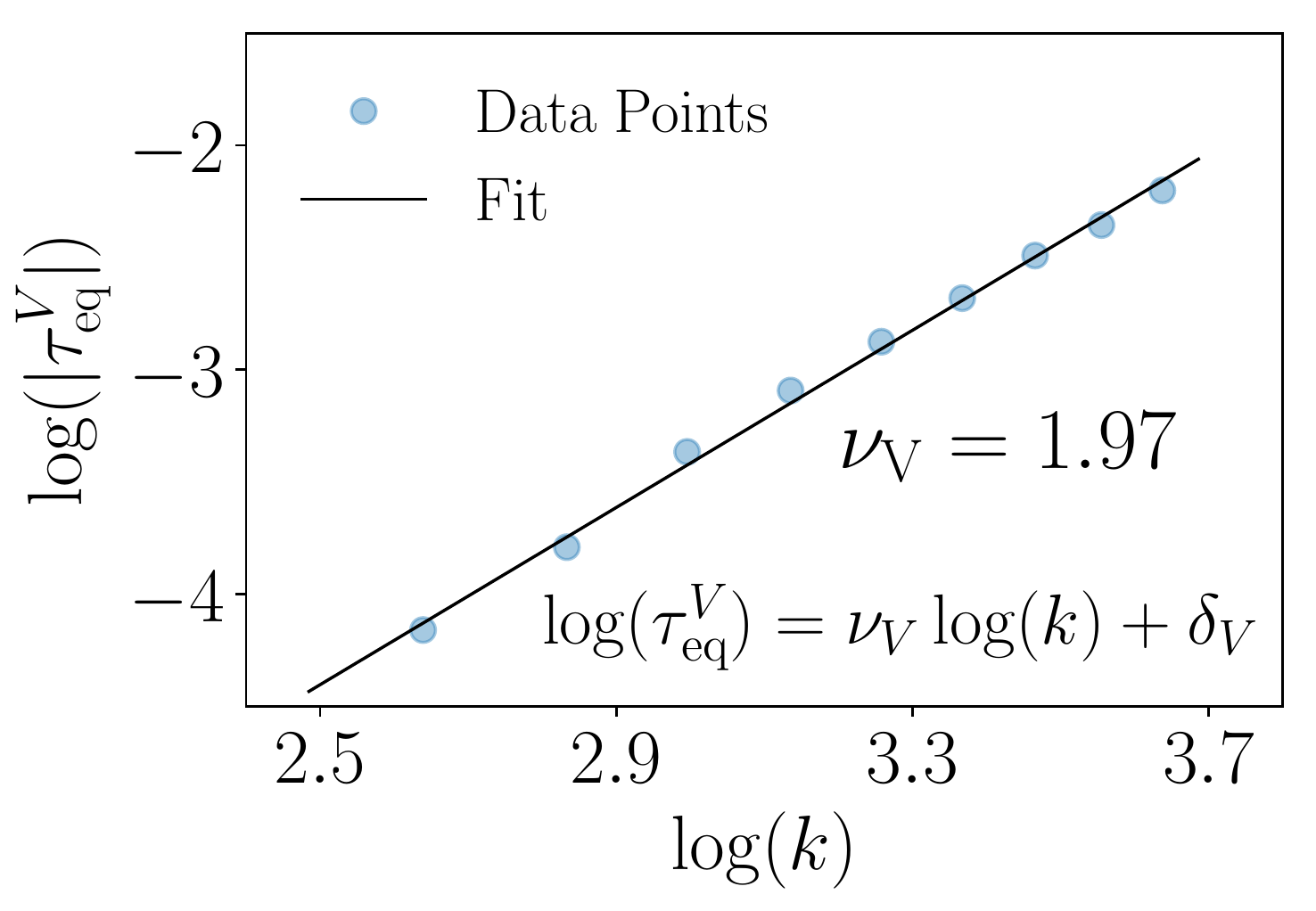}}
		\end{picture} 
	}	
	\includegraphics[width=0.5\textwidth]{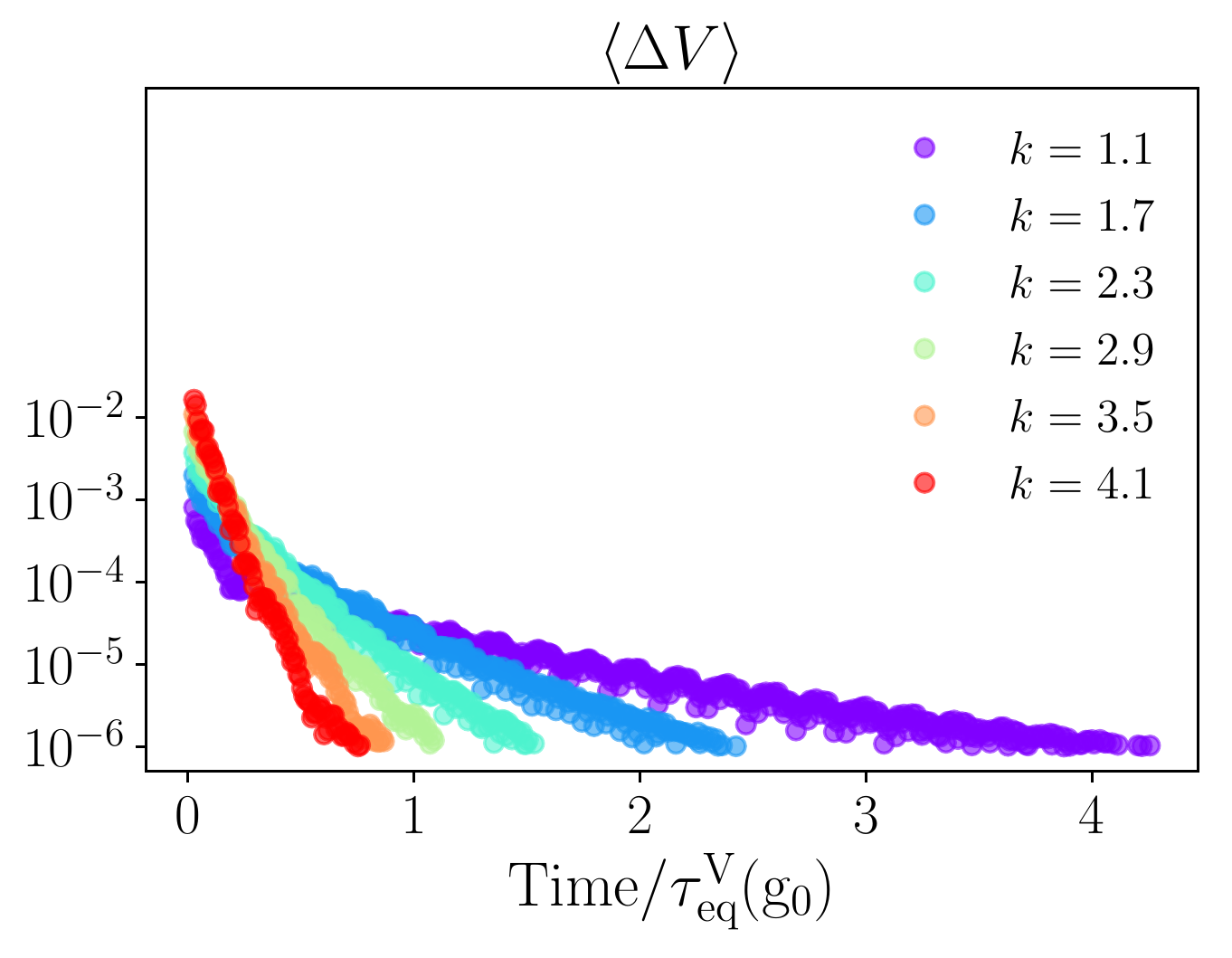}
\caption{{\bf Thermalization of the potential.} Moving average of the expectation value of 		the potential as a function of time for different coupling strengths. The value of $\left< \Delta V (t) \right>$ decays exponentially as a function of time. The slope becomes increasingly steeper for stronger coupling $g_{\rm max}(k)$, i.e. the thermalization is faster. 
{\bf Inset:} We extrapolate the decay of the relaxation time with a power law $\log( \tau^V_{\rm eq} (g) ) = \log(\tau^V_{\rm eq} (g_0)) - {\nu_{V}}\log(k)$. The optimal fit corresponds to $\nu_V = 1.97$ and $g_0^2 \tau^V_{\rm eq} (g_0) = 1.1$. Parameters: $N=300$, $g_0^2/\omega_S =0.01.$
}
\label{fig:pot_therm}
\end{figure}

\subsubsection{Thermalization in the CL model.}

We first study the dependence of the thermalization time on $g$ for observables on the system. 
In the simulation, we take as an initial state the thermal state of the non-interacting Hamiltonian $\rho^{\rm th}_{\beta}(t=0) = {\rho^{\rm th}_{\beta} (H^{(S)}) \otimes \rho^{\rm th}_{\beta} (H^{(B)})}$, and then perform a quench to a finite interaction strength $g = kg_0$, and consider the corresponding relaxation to the new equilibrium state.

In particular, in Fig.~\ref{fig:Hs_therm}, we compute the relative entropy $S(\rho \parallel \sigma) = \Tr \left( \rho \left(\log \rho - \log \sigma \right) \right)$ between the marginal of the time evolved state $\rho_{(S)} (t) = \Tr_B (\rho (t))$ and the thermal state of the system $\rho^{\rm th}_{(S)}(t) = \Tr_B[\rho_\beta^{\rm th}(H(t))]$. The relative entropy decays exponentially in time 
\begin{align}
S(\sys{\rho}(t) \parallel \rho^{\rm th}_{(S)} (t)) \sim e^{-t / \tau_{\rm eq} (g)},
\end{align}
where $\tau_{\rm eq} (g)$ is  the relaxation timescale for a given coupling strength $g$.
As we expect, the slope becomes increasingly steeper for stronger couplings. In order to understand the behaviour of $\tau_{\rm eq} (g)$ as a function of $g$, we assume a power law decay $\tau_{\rm eq} (g)  = \tau_{\rm eq} (g_0) {k}^{-\nu}$, where $\tau_{\rm eq} (g_0)$ corresponds to the relaxation time for $k = 1$ and $\nu$ quantifies the scaling with interaction strength. In Fig. \ref{fig:Hs_therm} (inset) we fit the function $\log (\tau_{\rm eq} (g) ) = \log(\tau_{\rm eq} (g_0)) - \nu \log(k)$ with a straight line, which confirms the scaling predicted by Eq.~\eqref{eq:scalingtau} with $\nu \approx 2$ even for rather large coupling strengths up to $g^2/\omega_S \approx 0.25$. 

Similarly, we need to verify that the interaction energy thermalizes and satisfies Eq.~\eqref{eq:scalingtau}. This is shown in Fig.~\ref{fig:pot_therm}, where we plot  $\Delta V = \langle V(t) \rangle  - \langle V_{\rm eq} \rangle$ for different values of the coupling $g_{\rm max}(k)$, and where $\langle V(t) \rangle$ is the exact value of the interaction energy for the unitary-evolved state 
 and $ V_{\rm eq}$ is its thermal equilibrium value (with respect to the global thermal state). By performing an extrapolation as the one of  Fig. \ref{fig:pot_therm}, we confirm the scaling in Eq. \eqref{eq:scalingThermalisation} up to the relatively large interaction strength of $g^2/\omega_S \approx 0.25$.

\subsubsection{Generalised covariance.} 

\begin{figure}
{\hspace*{-.5cm}
\includegraphics[width=0.5\textwidth]{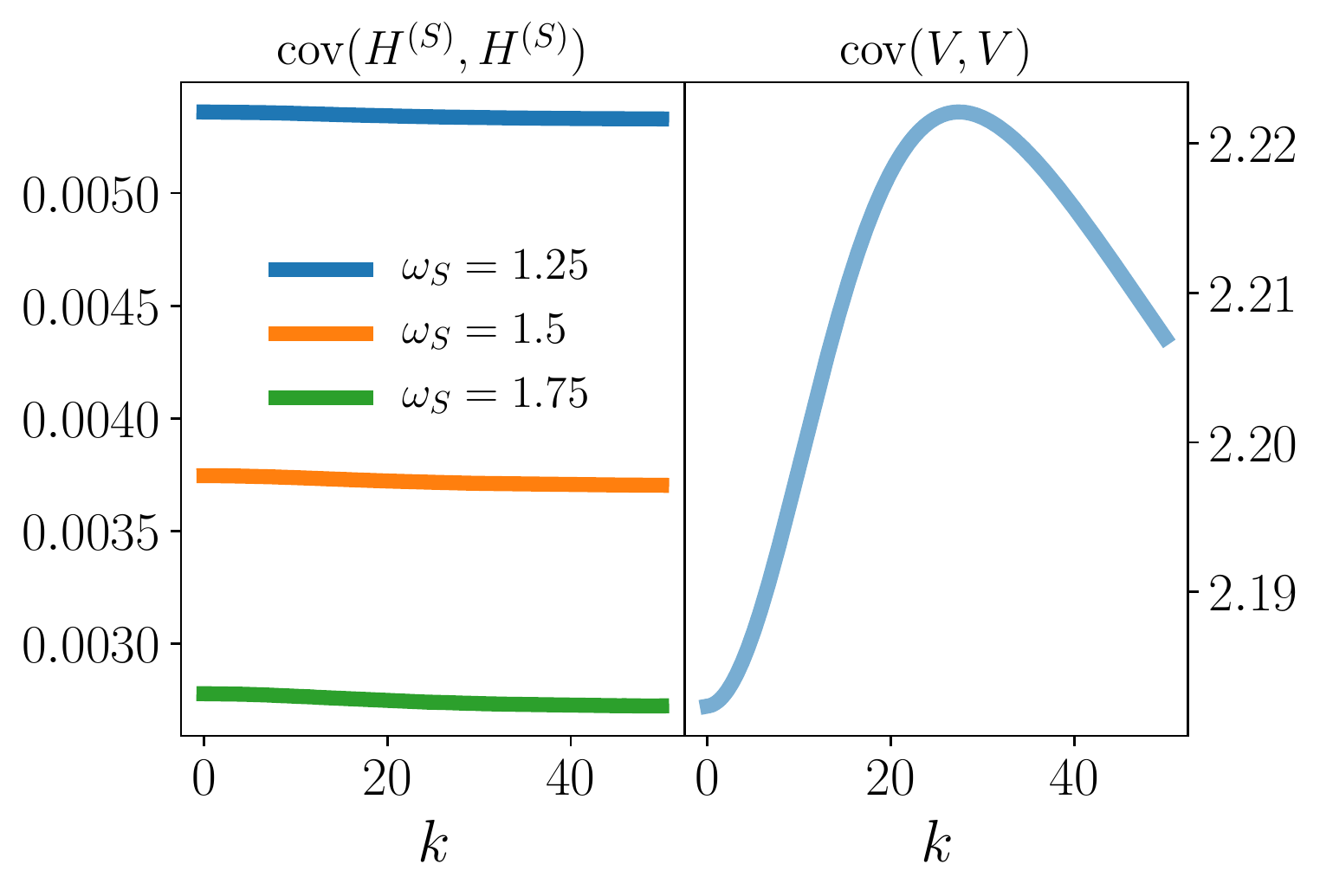}
\caption{{\bf Bounds on the dissipated work: covariance matrices.} In Eqs.~\eqref{eq:def_covariance} and \eqref{eq:Wdiss} we bounded the dissipated work by employing the covariance ${\rm cov}_{\omega_t} \left(  \tilde{H} , \tilde{H} \right)$. Notice that the covariance of $H_s$ and $V$ stays bounded for very large values of $k$.}
\label{fig:cov}
}
\end{figure} 

In Fig. \ref{fig:cov} we show the behavior of the covariance from Eq.~\eqref{eq:def_covariance} for the relevant quantities $H^{(S)}$ and $V$ as a function of the interaction strength $g$. One observes that ${\rm cov} (H^{(S)}, H^{(S)})$ stays essentially constant, which means that only $c^{(0)}$ in \eqref{expansionCov} contributes, hence also justifying Eq. \eqref{eq:choiceTiso}. On the other hand, ${\rm cov} (V, V)$ does vary with $g$, suggesting that higher order terms in the expansion in the expansion \eqref{expansionCov}  can play a role. 

\subsubsection{Speed-ups to isothermality}

\begin{figure}
	\includegraphics[width=0.5\textwidth]{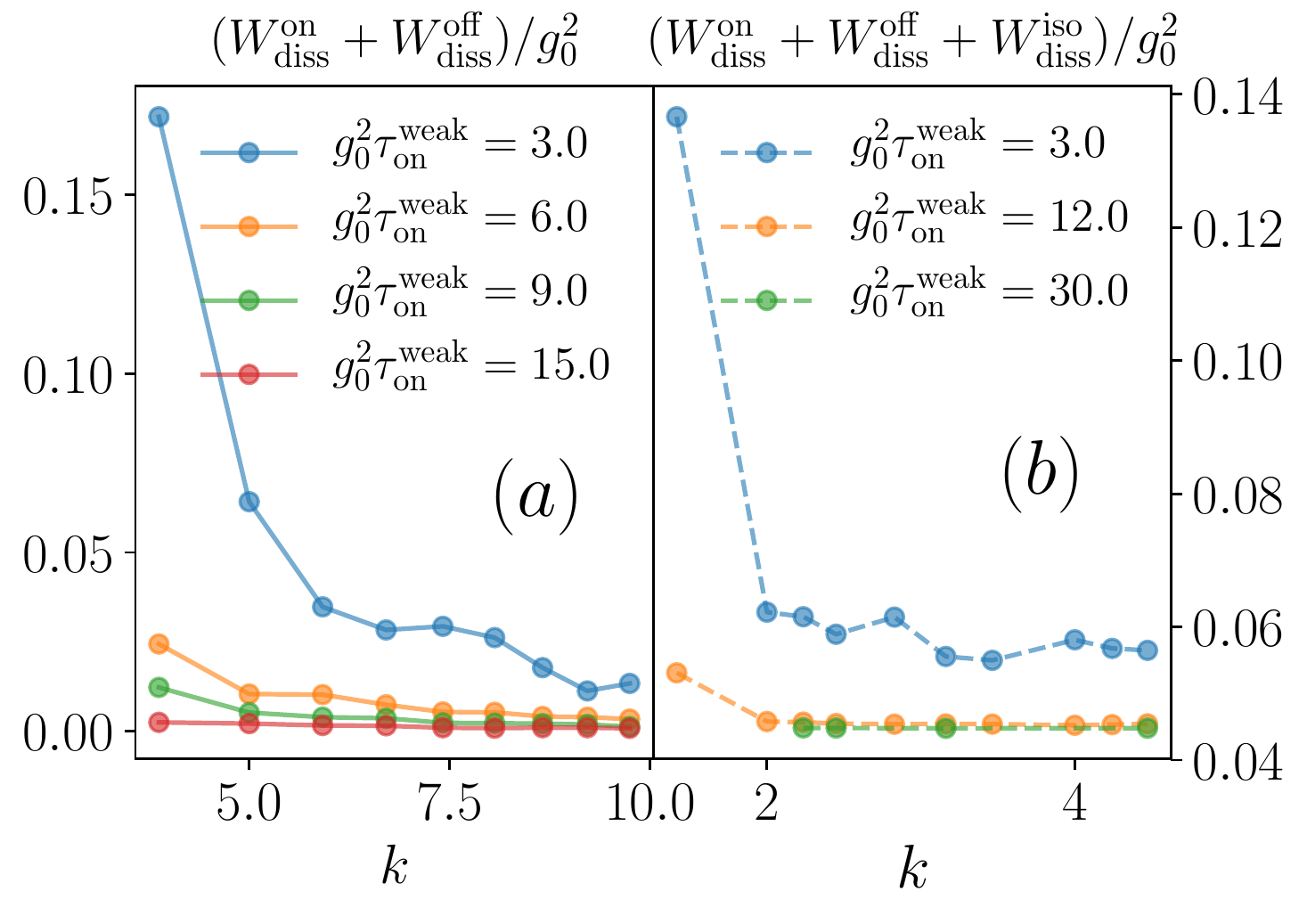}
	\caption{{\bf (a) Dissipated work for turning on and off the interaction in the CL model.}  For different values of $\tau^{\rm weak}_{\rm on}$ the error induced by going to the strong coupling regime decreases to zero. Parameters: $N = 300$, $g_0^2/\omega_S = 0.01 $ and $\beta\omega_S = 1.2$.
	{\bf (b)~Dissipated work for full protocols.} The total dissipation decays as  the interaction increases. For different values of $\tau^{\rm weak}_{\rm on}$, the error induced by going to the stronger coupling regime decreases to zero. The protocol consists of modifying the frequency of the systems from $\omega^i_S = \omega_S$ to $\omega^f_S = 2 \omega_S$. Parameters: $N = 300$,  $g_0^2/\omega_S = 0.01 $, $g_0^2\tau_{\rm iso}^{\rm weak} = 0.5$ and $\beta\omega_S = 1.2$.
	\label{fig:W_diss}}
\end{figure}
In order to confirm the intuition given by the generalised covariance, as a last step we simulate the full thermodynamic protocol and compute the associated dissipation. We vary $g$ by a linear ramp of the form given by Eq.~\eqref{eq:gpol} with $\alpha=1$. We take a very small initial coupling $g_i^2 = 10^{-8}\omega_S$ and a final coupling strength $g_f = k g_0$. Here, $g_0$ is a weak reference interaction strength that differs from $g_i$, unlike in Sec.~\ref{sec:speedups}. Nevertheless, as expected we find the same scaling of the dissipation with $g_f \propto k$.  

First, in Fig.~\ref{fig:W_diss} (a) we show the total dissipation $W_{\rm diss}$ for increasing and decreasing the interaction between system and bath while holding $\omega_S$ constant. The switching time is varied according to~$\tau_{\rm on} = k^2 \tau_{\rm on}^{\rm weak}$ in order to account for higher-order corrections to ${\rm cov}(V,V)$, as discussed in Sec.~\ref{sec:second_order}. One observes that $W_{\rm diss}^{\rm on/off}$ either decreases or stays constant with $k$, as expected from our analytic reasoning. Furthermore,  in Fig.~\ref{fig:W_diss} (b) we show the total dissipated work for the full thermodynamic protocol behaves in a similar way.  As we increase the interaction strength, the dissipation remains constant or drops close to zero, and as shown in Fig.~\ref{fig:TvsKk} the time substantially decreases. Hence, we have obtained the desired speed-ups. Regarding the timescales of the process, notice that the times shown in Fig.~\ref{fig:W_diss} are comparable to the thermalization times in Fig.~\ref{fig:Hs_therm} and \ref{fig:pot_therm}. This shows that the heuristic arguments of Sec.~\ref{sec:speedups} hold even for relatively fast driving.

\subsection{The resonant-level model}
\label{sec:resonant_level_model}

In this section, we benchmark our predictions using the analytically tractable resonant-level (RL) model. Specifically, the system of interest comprises a single distinguished fermionic mode coupled to an infinite collection of reservoir modes, also fermionic. The total system-bath Hamiltonian reads as $H = H^{(S)} + H^{(B)} + H^{(SB)}$, with 
\begin{align}
\label{eq:H_S_fermion}
H^{(S)} & = \varepsilon a^\dagger a,\\ 
\label{eq:H_B_fermion}
H^{(B)} & = \sum_k \omega_kb^\dagger_k b_k,\\
\label{eq:H_SB_fermion}
H^{(SB)} & = \sum_k \lambda_k\left (a^\dagger b_k + b^\dagger_k a\right ).
\end{align}
Here, $a$ annihilates a fermion with time-dependent energy $\varepsilon(t)$, while $b_k$ annihilates a fermion in the bath with energy $\omega_k$. We take coupling constants of the form $\lambda_k = g\bar{\lambda}_k$, where $g(t)$ is a time-dependent parameter characterising the overall interaction strength and the $\bar{\lambda}_k$ are time-independent weights. The relevant bath properties are characterised by the spectral density\footnote{Note that the spectral density is here defined to have units of frequency, in contrast to the CL model of Sec.~\ref{CLmodel} where customarily $\mathfrak{J}(\omega)\sim [{\rm frequency}]^2$.} $\mathfrak{J(\omega)} = 2\pi \sum_k  \lambda_k^2\delta(\omega- \omega_k)$, assumed to be of the form
\begin{equation}\label{eq:spectral_density_fermion}
\mathfrak{J}(\omega) = g^2\Theta(\Lambda-|\omega|),
\end{equation}
where $\Lambda$ is a high-frequency cutoff. 

\subsubsection{Solution for the dynamics}

Exact solutions for the RL model have recently been presented in the context of a debate regarding heat in strongly coupled open quantum systems, with particular emphasis on the wide-band limit $\Lambda\to \infty$~\cite{Ludovico2014, Esposito2015a, Esposito2015, Bruch2016, Haughian2018}.  Note, however, that the system-bath interaction energy is proportional to $\Lambda$, and thus formally divergent in this limit (this can be seen easily using the reaction-coordinate representation~\cite{Nazir2018}, for example). We thus take $\Lambda$ to be finite but much larger than all other energy scales. 

Under this assumption, we use a quantum Langevin approach to solve for the open-system evolution, detailed in Appendix~\ref{app:RLM}. Our approximate analysis requires that the dynamics proceeds much more slowly than the inverse cutoff scale $\Lambda^{-1}$, but otherwise allows for arbitrary driving protocols and strong system-bath coupling. Taking a factorised system-bath density matrix at the initial time, $\rho(0) = \rho_{(S)}(0) \rho^{\rm th}_\beta(H^{(B)})$, we find the level occupation $n(t) = \langle a^\dagger a\rangle$ and the system-bath interaction energy $v(t) = \langle H^{(SB)}\rangle$ to be given by
\begin{align}
\label{eq:level_occupation}
n(t) & = \frac{1}{2} + |K(t,0)|^2 \left(n(0) - \frac{1}{2}\right)\\
&\!\! \quad -  \frac{1}{2}\int_{0}^t {\rm d} s\!\int_{0}^t{\rm d }s' K(t,s) g(s) \phi(s-s')g(s') K^*(t,s'), \notag  \\
\label{eq:SB_correlation}
v(t) & = {\rm Im}\int_{0}^t{\rm d} s\, K^*(t,s) g(t)\phi(t-s) g(s).
\end{align}
These expressions are written in terms of the propagator
\begin{equation}\label{eq:propagator}
K(t,t') = \exp \left [\int_{t'}^t{\rm d} s \left ( -{\rm i}\varepsilon(s) - \frac{g(s)^2}{2}\right )\right ],
\end{equation}
and the noise correlation function
\begin{equation}\label{eq:noise_correlation_function}
\phi(t) = \frac{1}{{\rm i}\beta}\left [ \frac{1}{\sinh(\pi t/\beta)}  - \frac{\cos(\Lambda t)}{\pi t/\beta}\right ].
\end{equation}
Note that the second, cutoff-dependent contribution to $\phi(t)$ is essential to regulate the divergence of the integrand in Eq.~\eqref{eq:SB_correlation} as $s\to t$, but plays essentially no role in Eq.~\eqref{eq:level_occupation} for large~$\Lambda$. In Appendix~\ref{app:RLM} we show that the results for the dissipated work obtained within this approach converge to a $\Lambda$-independent value for sufficiently large $\Lambda$.

It follows immediately from Eq.~\eqref{eq:propagator} that the relaxation timescale is given by $\tau_{\rm eq} \sim 1/g^2$, in agreement with Eq.~\eqref{eq:scalingtau}, even though the evolution is non-Markovian, in general. Note also that, since Eq.~\eqref{eq:SB_correlation} contains one propagator while Eq.~\eqref{eq:level_occupation} includes two factors of $K(t,t')$, the relaxation timescale of the interaction energy is twice as long as that of $n(t)$. This is in accordance with the relaxation behaviour of the CL model shown in Figs.~\ref{fig:Hs_therm} and~\ref{fig:pot_therm}.

\subsubsection{Dissipated work}

\begin{figure}
	\includegraphics[width=0.45\textwidth]{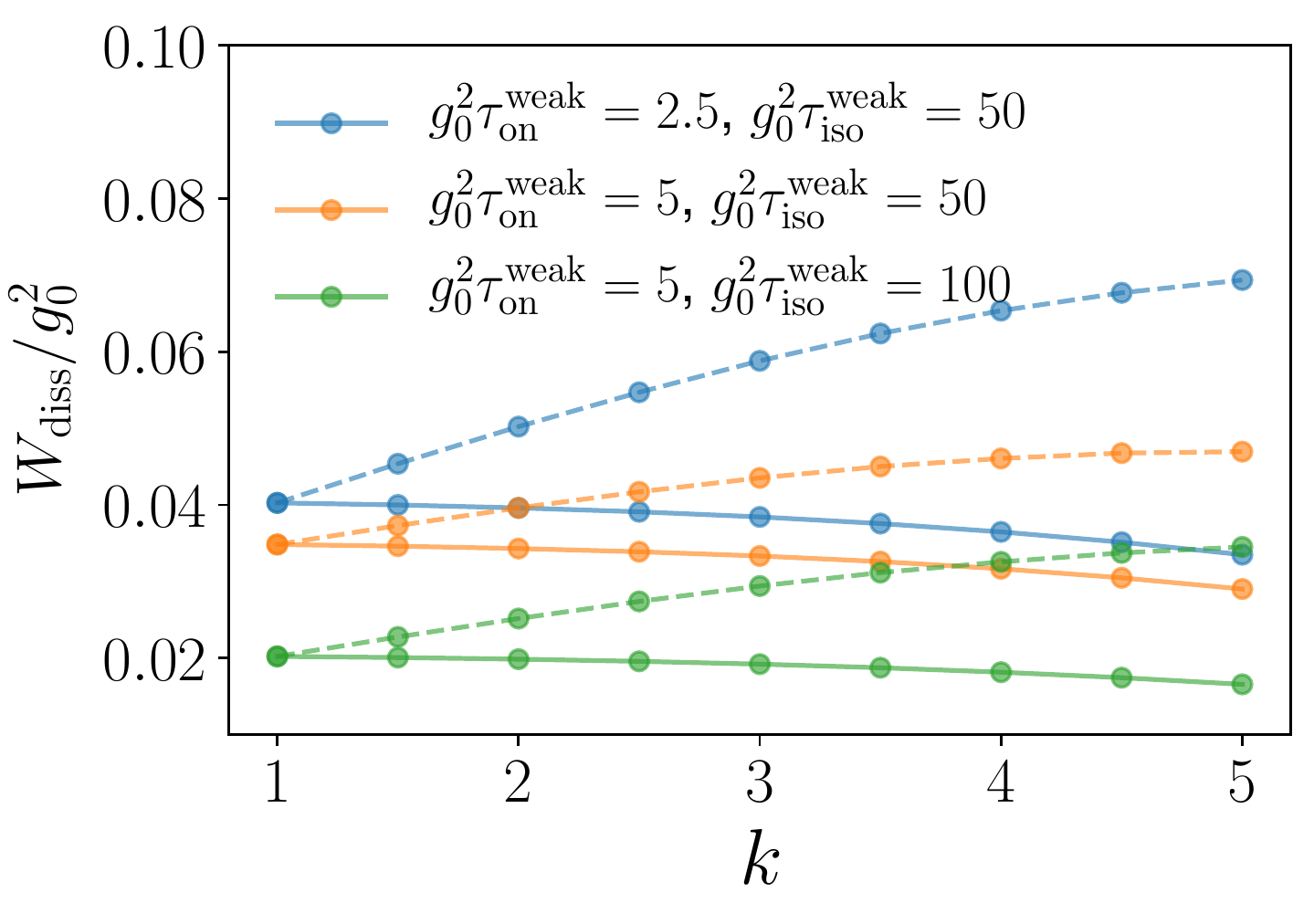}
    \caption{Dissipated work $W_{\rm diss}$ over an entire protocol in the RL model for various switching times and values of $k$. As a function of $k$, the isotherm time is reduced as $\tau_{\rm iso} = \tau^{\rm weak}_{\rm iso}/k^2$, while the switch-on and switch-off times are increased either as $\tau_{\rm on} = k\tau^{\rm weak}_{\rm on}$ (dashed lines) or $\tau_{\rm on}  = k^2\tau^{\rm weak}_{\rm on}$ (solid lines). Parameters: $\varepsilon_f = 2\varepsilon_i$, $g_0^2/\varepsilon_i =  0.1$,  $\beta\varepsilon_i = 1$ and $\Lambda/\varepsilon_i = 100$. \label{fig:W_diss_RL}}
\end{figure}

We now compute the dissipated work during an isothermal protocol, where the level energy $\varepsilon(t)$ is linearly ramped from an initial to a final value $\varepsilon_i \to \varepsilon_f$ while interacting with the bath. The dissipated work is given by $W_{\rm diss} = W - \Delta F$, where $W$ is found from Eqs.~\eqref{work_def},~\eqref{eq:level_occupation} and~\eqref{eq:SB_correlation}, while $\Delta F = \varepsilon_f - \varepsilon_i + \beta^{-1} \ln \left [f(\varepsilon_f)/f(\varepsilon_i)\right ]$, with the Fermi-Dirac distribution $f(\omega) = (\ee^{\beta\omega}+1)^{-1}$. At the start and end of the protocol, the system-bath interaction energy is switched on according to Eq.~\eqref{eq:gpol} with $\alpha = 1$, $g_i = 0$ and $g_f = kg_0$, and switched off via the reverse procedure. Note that here again we have $g_0\neq g_i$, unlike in Sec.~\ref{sec:speedups}, yet as expected we  find a similar behaviour  of dissipation with the final coupling strength $g_f\propto k$.

To confirm this scaling, we plot $W_{\rm diss}$ for several different parameters as a function of $k$ in Fig.~\ref{fig:W_diss_RL}. We see that the dissipation grows sublinearly with $k$ for $\tau_{\rm on} =  k\tau_{\rm on}^{\rm weak}$, while the dissipation strictly decreases for the more conservative choice of $\tau_{\rm on} =  k^2\tau_{\rm on}^{\rm weak}$. This suggests that, as in the Caldeira-Leggett model, the generalised covariance ${\rm cov}(V,V)$ does depend on $g$, necessitating higher-order terms in the expansion~\eqref{expansionCov} to be taken into account. Nevertheless, the results confirm that control over the system-bath interaction can indeed reduce the time taken by an isothermal process without incurring additional dissipation (c.f.~Fig.~\ref{fig:TvsKk} showing the time of the isothermal process).

\begin{figure}
    \centering
    \includegraphics[width = \linewidth]{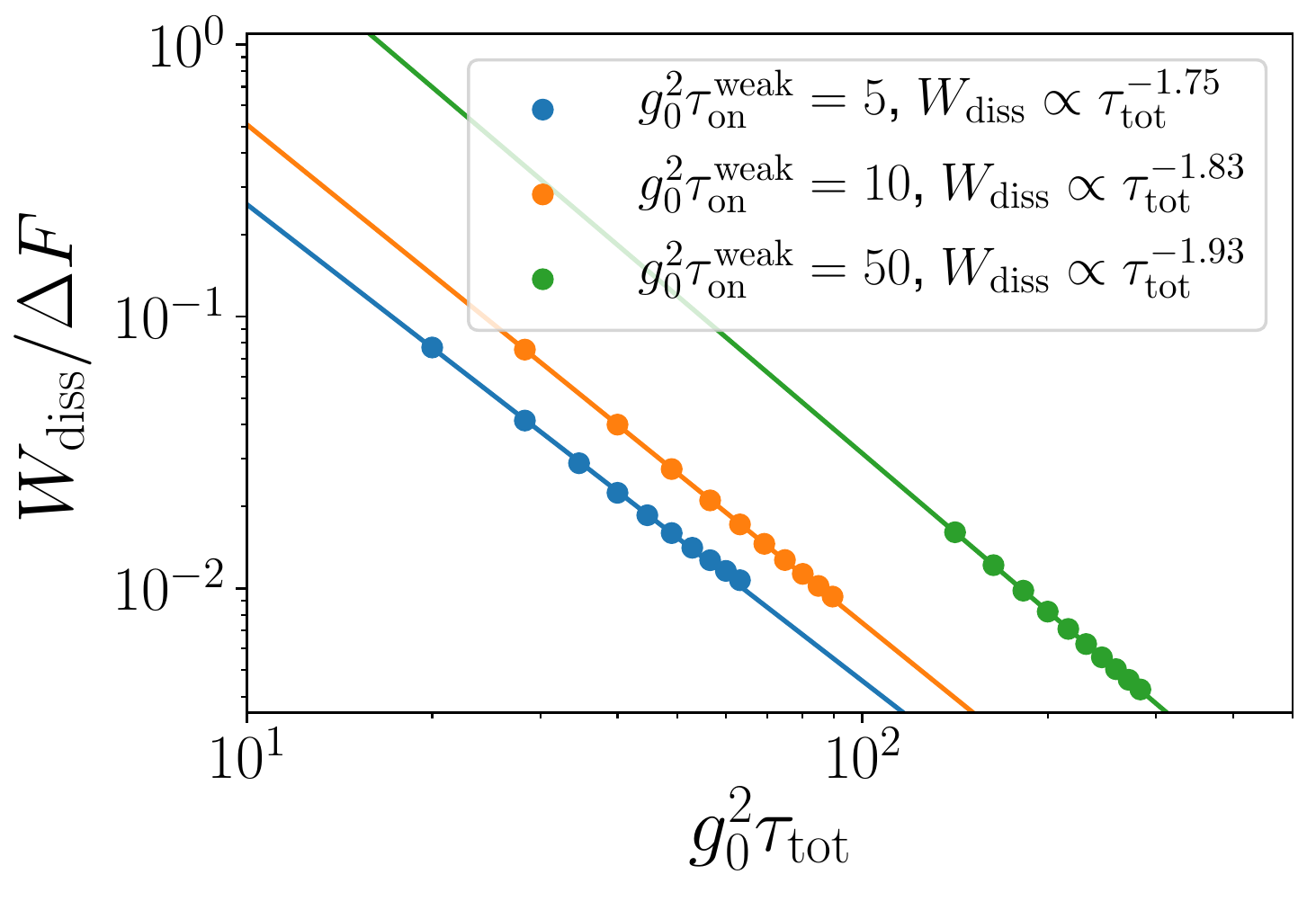}
    \caption{Decay of the optimal dissipated work as a function of the total protocol time in the RL model. The optimal $k$ for each $\tau_{\rm tot}$ is found following the procedure discussed below Eq.~\eqref{eq:tau_tot_ksqd}. Circles show numerical results, lines indicate power-law fits to the function $W_{\rm diss}(\tau_{\rm tot}) = C \tau_{\rm tot}^\nu$. Parameters: $\varepsilon_f = 2\varepsilon_i$, $g_0^2/\varepsilon_i = 0.1$, $\beta\varepsilon_i = 1$, $\Lambda/\varepsilon_i = 100$.}
    \label{fig:W_decay}
\end{figure}

We now consider how the dissipated work scales with the total time of the protocol in the optimal case, as discussed in Sec.~\ref{sec:optimalscaling}. In particular, we focus on protocols where $\tau_{\rm on}$ is proportional to $k^2$ in order to ensure that the dissipated work is non-increasing as $k$ increases. Following the procedure outlined below Eq.~\eqref{eq:tau_tot_ksqd}, we compute the optimal $W_{\rm diss}$ as a function of $\tau_{\rm tot}$, holding $\tau_{\rm on}^{\rm weak}$ fixed. The results are plotted in Fig.~\ref{fig:W_decay}, showing that the dissipated work decays as a power law,  $W_{\rm diss} \propto \tau_{\rm tot}^{-\nu}$, to a good approximation over the range of times considered. As the time to turn the system-bath interaction on and off is increased, the power-law exponent $\nu\to 2$; as predicted in Sec.~\ref{sec:optimalscaling}. Faster switching of the interaction incurs additional dissipation which was not accounted for in Eq.~\eqref{eq:Wdisweakk}, thus leading to smaller exponents $\nu<2$. Alternatively, one could also obtain $\nu \approx 2$  by reducing $g_0$, i.e., by working in the weak coupling regime (in our simulations $g_0^2=0.1\varepsilon_i$ which is  certainly non-negligible). It is also important to note that a better \textit{scaling} of $W_{\rm diss}$ does not necessarily correspond to less overall dissipation. Indeed, for the parameter regime considered in Fig.~\ref{fig:W_decay}, $W_{\rm diss}$ for a given $\tau_{\rm tot}$ is minimised by choosing a smaller value of $\tau_{\rm on}^{\rm weak}$, since this allows more time for the isothermal part of the process to take place slowly. 

\begin{figure}
    \centering
    \includegraphics[width=0.95\linewidth]{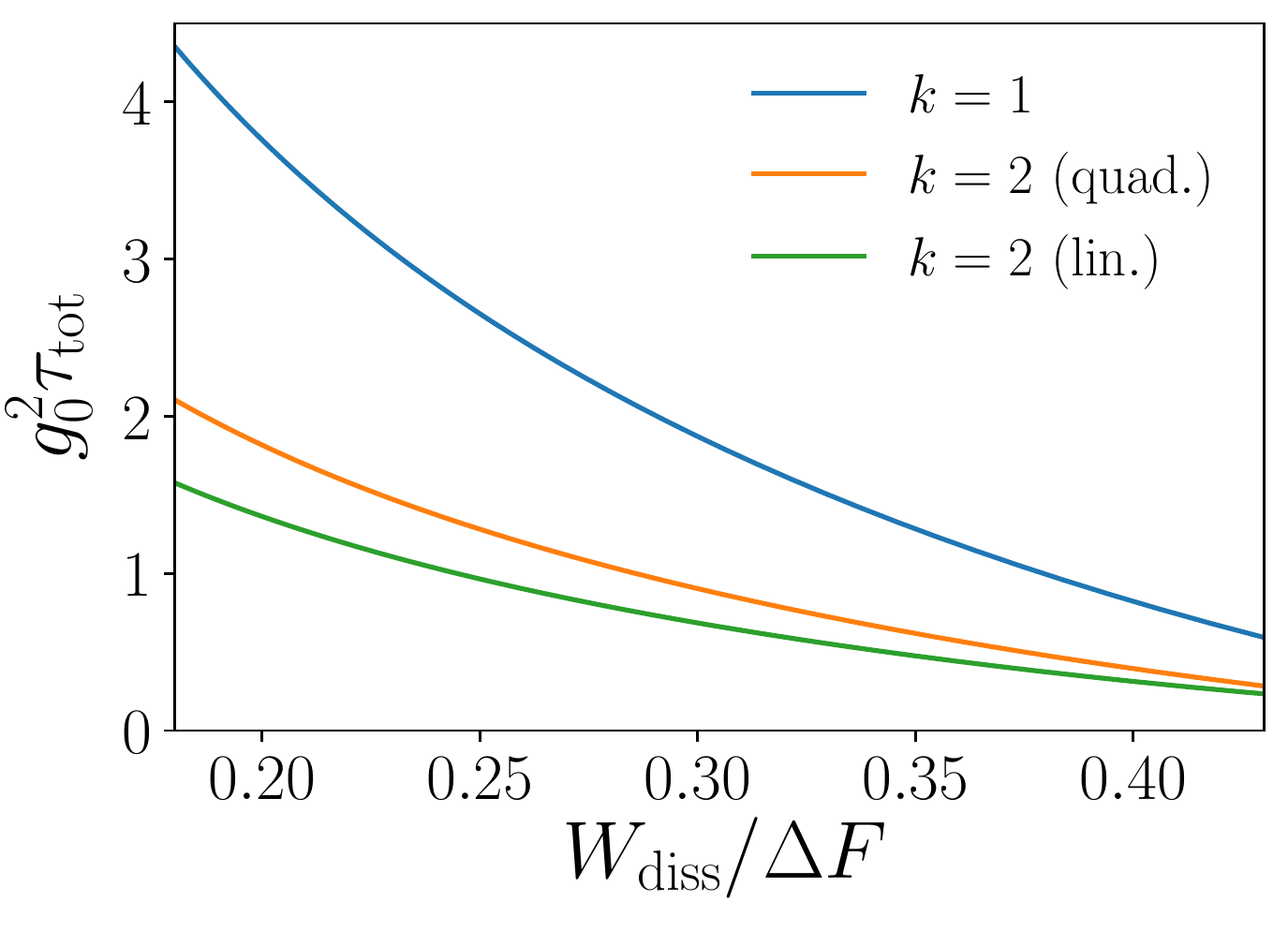}
    \caption{Total protocol time as a function of dissipated work in the RL model. Each curve is generated by varying $g_0^2\tau_{\rm tot}^{\rm weak}\in [0.5, 5]$ while holding $\tau_{\rm iso}^{\rm weak}/\tau_{\rm on}^{\rm weak} = 30$ fixed. The curves for~$k=2$~show both linear scaling, $\tau_{\rm on} = k\tau_{\rm on}^{\rm weak}$, and quadratic scaling, $\tau_{\rm on} = k^2\tau_{\rm on}^{\rm weak}$. Parameters: $\varepsilon_f = 2\varepsilon_i$, $g_0^2/\varepsilon_i = 0.01$, $\beta\varepsilon_i = 1.1$ and $\Lambda/\varepsilon_i=100$.}
    \label{fig:work_ttot}
\end{figure}
So far in this section we have focussed on the regime of slow driving where $g_0^2\tau_{\rm tot} \gg 1$. We now show that our approach also works in the complementary fast-driving regime. In Fig.~\ref{fig:work_ttot} we plot the total time versus dissipated work, comparing sped-up protocols with $k=2$ to reference protocols with $k=1$. For a given dissipation, the SI significantly reduces the total time taken for the isothermal transformation, even when $g^2_0\tau_{\rm tot} < 1$. This represents further evidence that our proposed SI work well outside of the slow-driving regime assumed in Eq.~\eqref{eq:Wdiss}.

\subsubsection{Carnot cycle}

\begin{figure}
    \centering
    \includegraphics[width=0.95\linewidth]{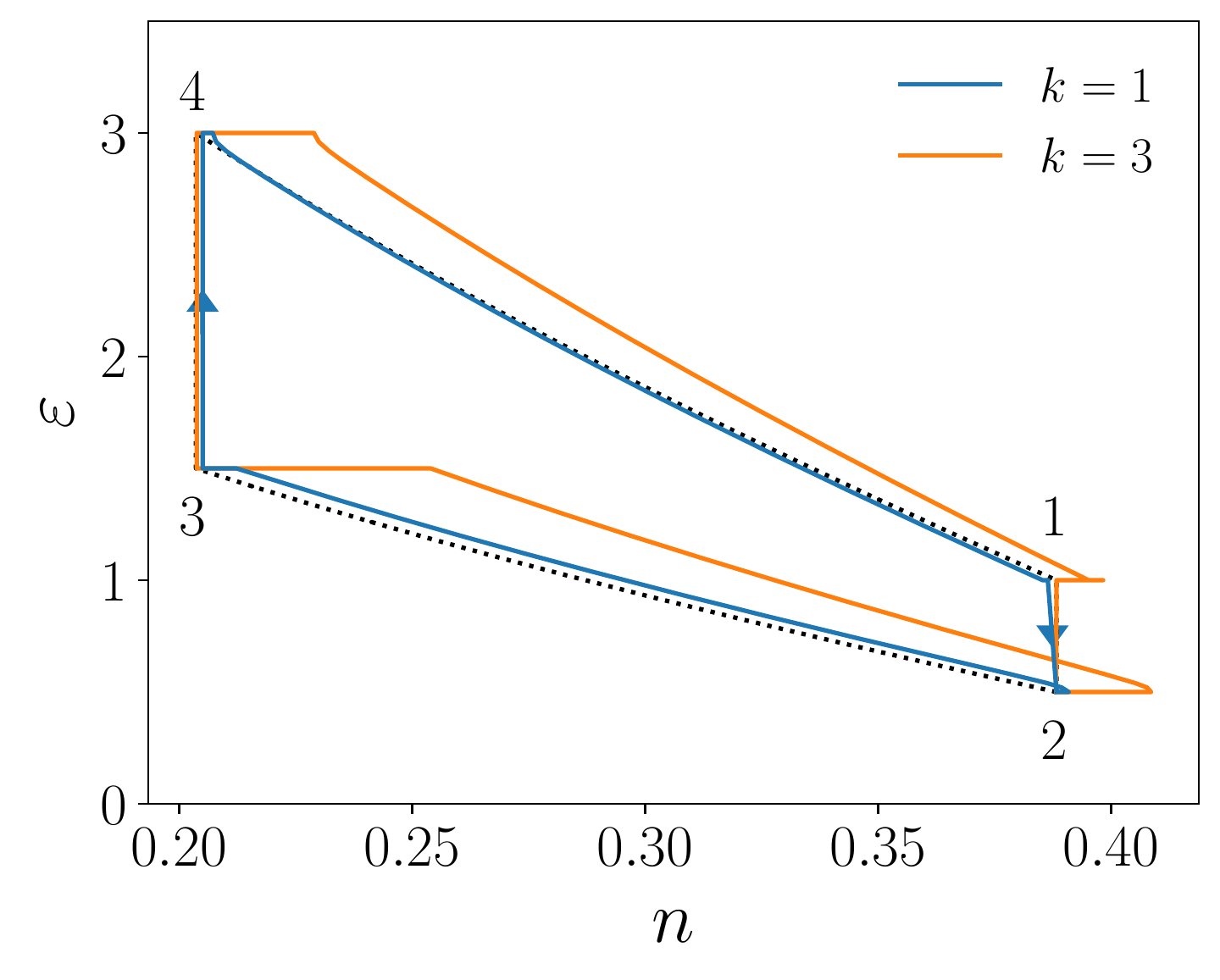}
    \caption{Carnot cycle for the RL model showing the level's energy $\varepsilon(t)$ versus its occupation $n(t)$. For finite system-bath interaction strengths (blue and orange solid lines), correlations with the bath lead the system state during the isotherms to deviate significantly from equilibrium (dotted line). The protocol is the same as Fig.~\ref{fig:Carnot_lin_sqd}(a) with $g_0^2\tau_{\rm on}^{\rm weak}~=~25$.}
    \label{fig:Carnot_cycle}
\end{figure}

\begin{figure*}
    \begin{minipage}{0.48\linewidth}
    \centering
   \includegraphics[width=\linewidth]{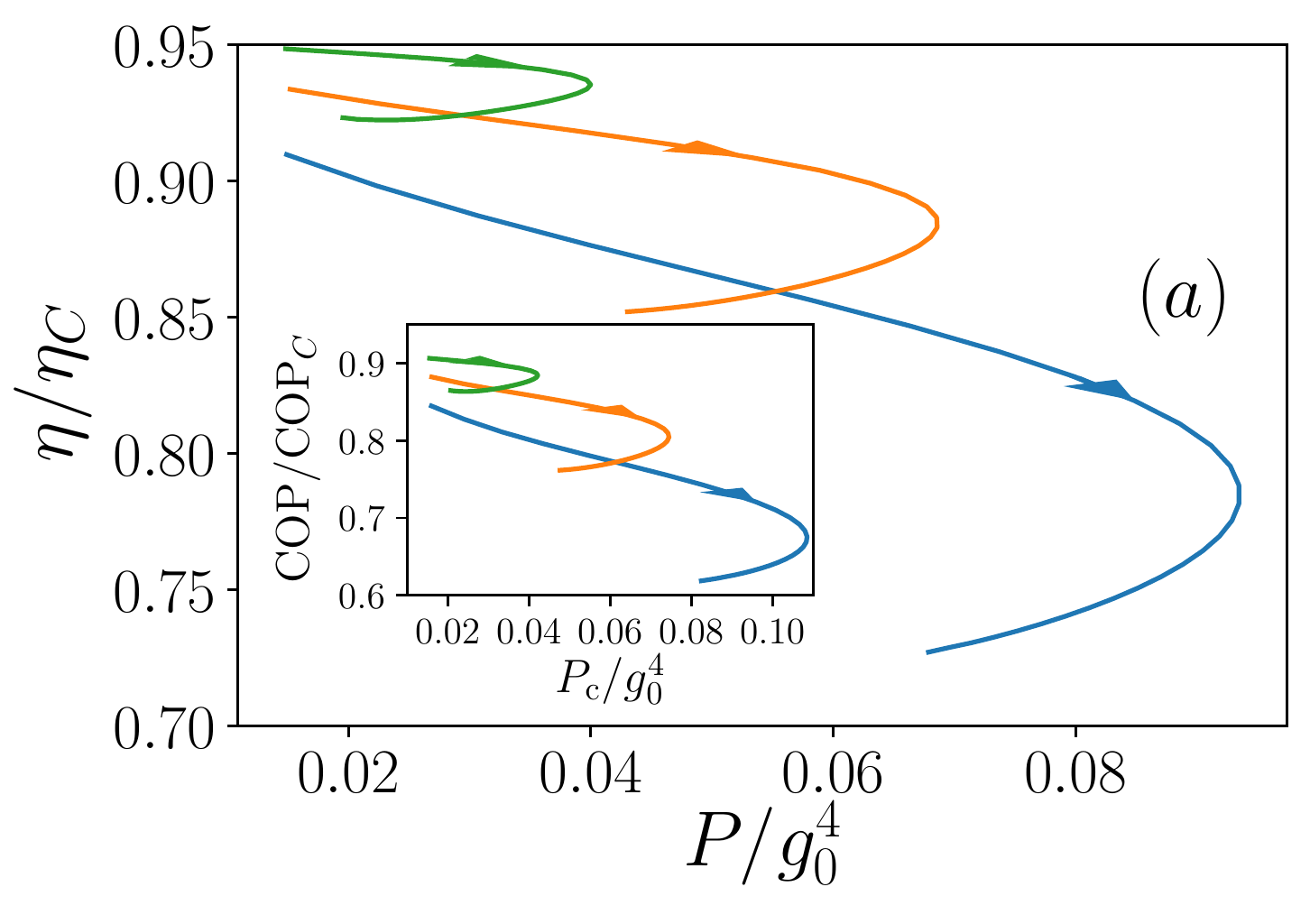}
   \end{minipage}
    \begin{minipage}{0.48\linewidth}
    \centering
     \includegraphics[width=\linewidth]{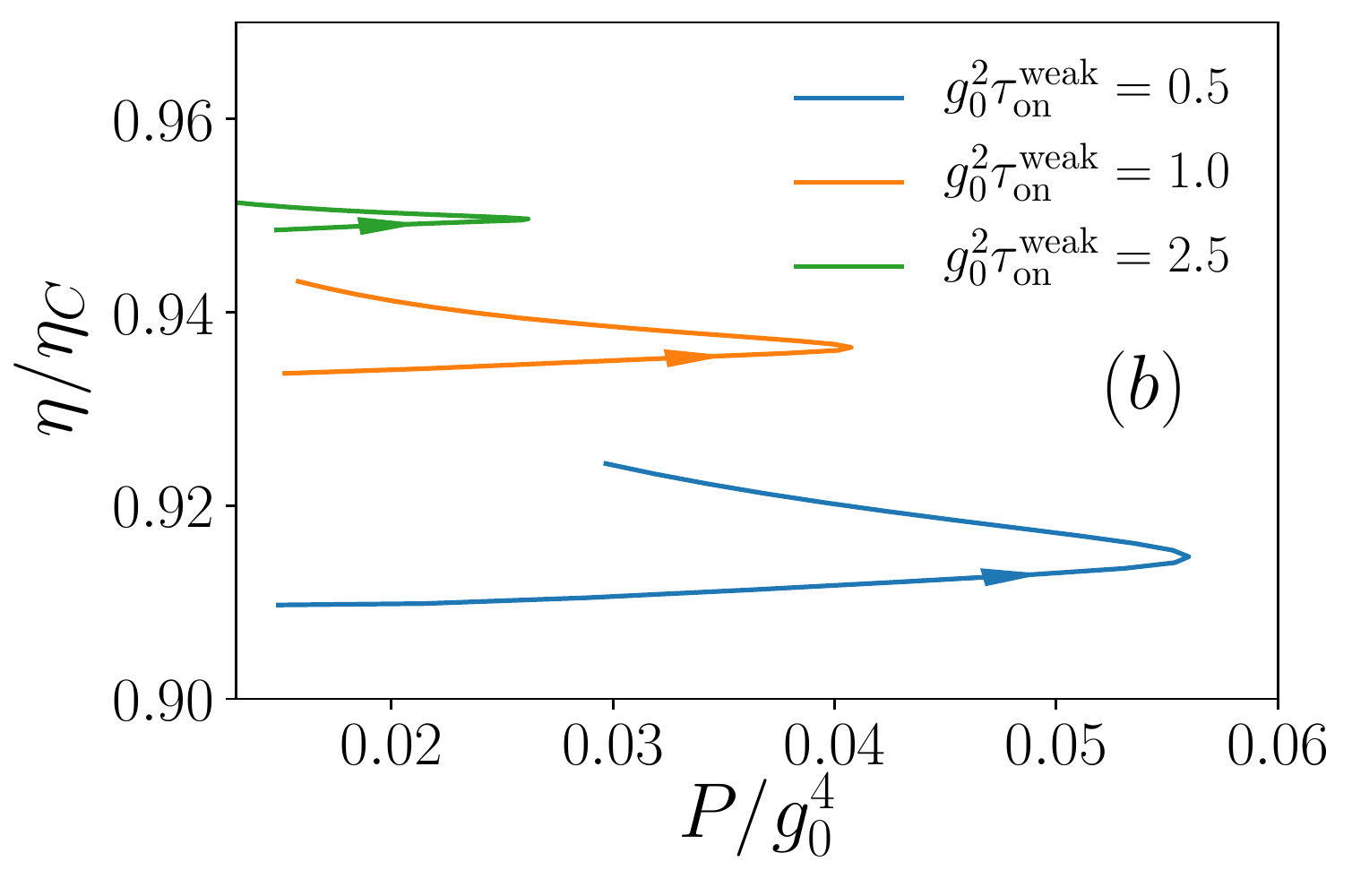}
    \end{minipage}
    \caption{Efficiency versus power for the quantum Carnot cycle depicted in Fig.~\ref{fig:Carnot_cycle}. Each curve shows engine performance for a fixed value of $\tau_{\rm on}^{\rm weak}$ and $\tau_{\rm iso}^{\rm weak}$ and various values of $k\in [1,8]$, with arrows indicating increasing $k$. The switching on and off of the interaction is scaled differently in the two plots, with (a) $\tau_{\rm on} = k\tau_{\rm on}^{\rm weak}$ and (b) $\tau_{\rm on} = k^2\tau_{\rm on}^{\rm weak}$. Parameters: $[\varepsilon_1,\varepsilon_2,\varepsilon_3,\varepsilon_4]=[1,0.5,1.5,3]$, $g_0^2/\varepsilon_1 = 0.1$, $g_0^2\tau_{\rm iso}^{\rm weak} = 50$, $(\beta_c\varepsilon_1)^{-1} = 1.1$, $(\beta_h\varepsilon_1)^{-1} = 2.2$ and $\Lambda/\varepsilon_1 = 100$. The inset of (a) shows a near-identical refrigerator performance characteristic for the reverse cycle. \label{fig:Carnot_lin_sqd}}
\end{figure*}

As a final demonstration of our speed-ups to isothermality, we show that the power of a quantum thermal machine can be improved without sacrificing its efficiency. We consider the Carnot-like engine cycle indicated in Fig.~\ref{fig:Carnot_cycle}, which proceeds by the following steps: (i)~adiabatic expansion $\varepsilon_1 \to \varepsilon_2$, (ii)~isothermal compression $\varepsilon_2 \to \varepsilon_3$ in contact with a cold bath at inverse temperature $\beta_c$, (iii)~adiabatic compression $\varepsilon_3 \to \varepsilon_4$, (iv)~a final isothermal expansion $\varepsilon_3 \to \varepsilon_4$ in contact with a hot bath at inverse temperature $\beta_h$. The density matrix of the system mode commutes with the Hamiltonian at all times during the adiabatic steps. It is therefore possible to perform the adiabatic steps arbitrarily quickly without affecting the rest of the protocol, and in the following we assume that the adiabats are instantaneous. 

As in the previous subsection, we consider isothermal protocols of the form of Eq.~\eqref{eq:gpol} with $\alpha = 1$, $g_i=0$ and $g_f = kg_0$. We focus on cycles where $\varepsilon_2/\varepsilon_1  = \varepsilon_3/\varepsilon_4 = \beta_h/\beta_c$. This choice ensures that the system is close to equilibrium with the new bath temperature at the start of each isotherm, thus minimising the dissipation incurred by switching on the coupling to the bath. We use Eqs.~\eqref{eq:level_occupation} and \eqref{eq:SB_correlation} to describe the evolution during the hot and cold isotherms, assuming that the corresponding bath relaxes back to equilibrium over the course of the subsequent isotherm. Since the working medium is pushed far from equilibrium during engine operation, we need to repeat the engine cycle several times until a limit cycle is reached. In our calculations, the cycle is repeated until the initial and final level occupation differ by less than~1\%.

For a given $\tau_{\rm on}^{\rm weak}$ and $\tau_{\rm iso}^{\rm weak}$, we study how the power and efficiency of an engine cycle scales with $k$. We consider two different scenarios, as shown in Fig.~\ref{fig:Carnot_lin_sqd}. By scaling the switch-on and switch-off time as $\tau_{\rm on}\propto k$, we obtain a large improvement in power due to the significant reduction in the total cycle time. However, this comes at the cost of losing some efficiency because the work dissipated during each isotherm increases with $k$ (see Fig.~\ref{fig:W_diss}(c)). On the other hand, if we instead use the more conservative choice $\tau_{\rm on}\propto k^2$, we find that \textit{both} power and efficiency can be improved by increasing $k$, since both the total time of the protocol and the dissipation decrease. However, this more conservative scaling naturally corresponds to smaller enhancements of power.

One may also realise a refrigerator by operating the Carnot cycle in the opposite direction. As an example, the inset of Fig.~\ref{fig:Carnot_lin_sqd}(a) shows the coefficient of performance as a function of cooling power for the reverse cycle. We find very similar qualitative characteristics to the corresponding engine. Our approach can therefore also be used to boost the power of Carnot-like refrigerator cycles while retaining efficient performance.

\section{Experimental feasibility and robustness\label{sec:feasibility}}

The implementation of our proposal requires the ability to smoothly modulate both the Hamiltonian of the system and its coupling to the environment. In this section, we discuss two experimental platforms where such control is feasible. We also show that our protocol is robust against unavoidable control errors.

\subsection{Impurities in cold atomic gases}

A promising candidate system is a cold atomic gas with impurity atoms of another species immersed within it.  Such binary mixtures of cold atoms have been studied in numerous experiments in recent years~\cite{Silber2005, Spiegelhalder2009,Catani2009, Catani2012,FerrierBarbut2014,Cetina2015, Hohmann2016, Hohmann2017, Lous2017, Bouton2020}. The impurities behave as controllable open quantum systems interacting with their ultracold gas  environment~\cite{Recati2005,Klein2007,Goold2011,Haikka2013,Hangleiter2015,Mitchison2016,Lampo2017,Cosco2018}. Examples of useful thermodynamic cycles in this context include refrigeration of the impurities~\cite{Griessner2006} or of the surrounding gas~\cite{Niedenzu2019}.

The control required for our scheme may be implemented by confining the impurities by a species-selective dipole potential that can be dynamically modulated~\cite{Catani2009,Catani2012,Lous2017}. Crucially, moreover, the system-bath coupling can be controlled by tuning the $s$-wave scattering length describing interspecies collisions via a Feshbach resonance~\cite{Chin2010}. This allows the scattering length to be gradually~\cite{Haller2009} or suddenly~\cite{Cetina2015} varied over several orders of magnitude or even set to zero without affecting the environment properties~\cite{Catani2012}.

To be concrete, let us consider a setup similar to the experiments reported in Ref.~\cite{Catani2012}, where an impurity atom is embedded in a Bose-Einstein condensate (BEC) of a different species. The open system is thus a harmonic oscillator corresponding to vibrations of the trapped impurity, which is damped by collisions with the BEC atoms. Phonon excitations in the BEC behave like a bosonic bath~\cite{Recati2005,Klein2007,Haikka2013,Hangleiter2015} and the impurity-BEC system can be described by the CL model of Sec.~\ref{CLmodel} if the dynamics of the condensate mode is neglected~\cite{Johnson2012,Lampo2017}. Translating the experimental parameters of Ref.~\cite{Catani2012} into our notation gives an impurity trapping frequency of $\omega_S = 2\pi\times 1.0~$kHz, a temperature of $\beta\omega_S\approx0.02$ and measured damping rates on the order of $\tau_{\rm eq}(g) \omega_S \gtrsim 10$. All of the ingredients necessary for the implementation of SI have therefore already been demonstrated in the context of ultracold gases.

\subsection{Semiconductor quantum dots}

Another potential platform to realise our scheme is a mesoscopic electronic device, such as a quantum dot or metallic island connected to a conducting electron reservoir~\cite{Hanson2007}. Here, the charge state localised on the dot or island exchanges particles and energy with the reservoir via tunnelling processes. This is similar to the RL model considered in Sec.~\ref{sec:resonant_level_model}, albeit with an additional feature: the Coulomb interaction typically plays an important role in mesoscopic electronics. Nonetheless, our general arguments still apply to these systems. Numerous thermodynamic protocols have already been experimentally implemented in this context, including a Szilard engine~\cite{Koski2014a,Koski2014b}, a refrigerator~\cite{Koski2015} and an autonomous heat engine~\cite{Josefsson_2018}.

Both the dot's energy level and the tunnel barriers that define its coupling to the reservoir can be independently~\cite{Ciorga2000,Elzerman2003} and dynamically~\cite{Koski2014a,Koski2014b} tuned by applying appropriate gate voltages to different parts of the system. Control over the tunnelling rate spanning several orders of magnitude has been demonstrated~\cite{Rochette2019}. Typical experimental parameters can be estimated from Ref.~\cite{Josefsson_2018}, which reports dot energies relative to the chemical potential on the order of $\varepsilon\lesssim 1~$meV, comparable temperatures of $\beta\varepsilon \gtrsim 1$ and tunnelling rates on the order of $g^2 \sim 10~{\rm GHz}\sim 0.01~$meV. Therefore, mesoscopic electronic devices seem equally promising for the implementation of SI.

\subsection{Robustness against error}

\begin{figure}
    \centering
    \includegraphics[width=0.95\linewidth]{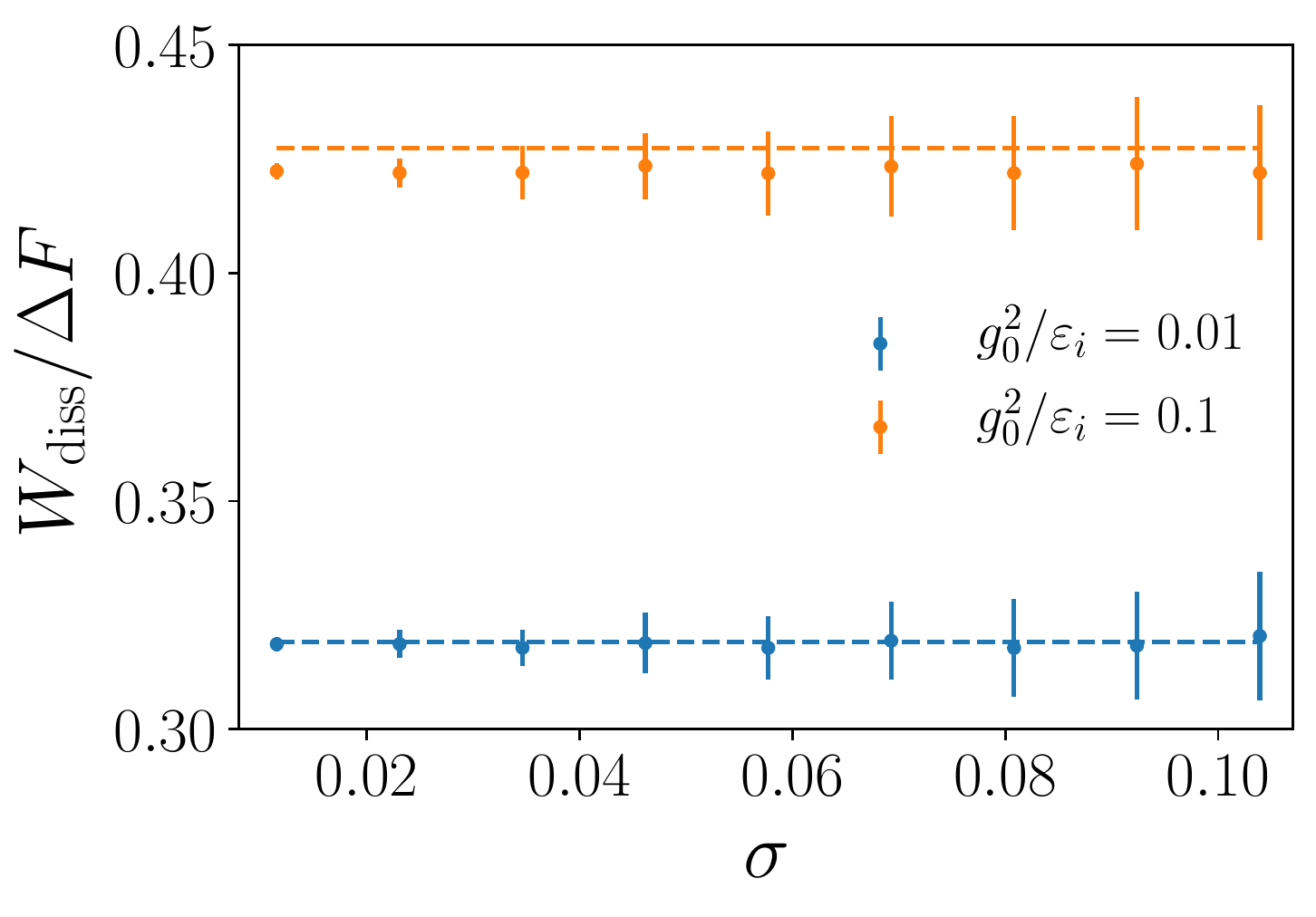}
    \caption{Dissipated work for noisy protocols in the RL model. Each data point represents 100 realisations of a random protocol with relative error $\sigma$ in the timing of each stroke (see main text for details). The points show the mean while the error bars show the variance of the dissipated work for a noisy protocol with $k=2$ and quadratic scaling of the switch-on and switch-off times: $\overline{\tau_{\rm on}} = \overline{\tau_{\rm off}} = k^2\tau_{\rm on}^{\rm weak}$ and $\overline{\tau_{\rm iso}} =\tau_{\rm iso}^{\rm weak}/k^2$. Dotted lines show the work done for noiseless reference protocols with $k=1$. Parameters: $\varepsilon_f= 2\varepsilon_i$, $\beta\varepsilon_i = 1.1$, $g_0^2\tau_{\rm on,off}^{\rm weak} = 0.15$, $g_0^2\tau_{\rm iso}^{\rm weak} = 1.5$ and $\Lambda/\varepsilon_i=100$.}
    \label{fig:schedule_noise}
\end{figure}
Any real experimental implementation suffers from unavoidable noise and control errors. It is therefore crucial to ensure that our scheme is robust against such imperfections. Assuming that the thermal bath represents the dominant source of dissipation and decoherence in the system, the key remaining issue is the extent to which the SI is affected by fluctuations in the applied control fields.

Since the analytical arguments of Sec.~\ref{sec:speedups} rely on perturbative arguments and were shown to hold for a broad family of protocols, we expect the SI to remain robust under small errors that do not lead to large and discontinuous changes in the control fields. In order to quantitatively demonstrate this, we consider a specific yet realistic kind of noise: namely, imperfect timing of the control operations. In particular, we assume that the duration of each step of the protocol is given by $\tau_\alpha = \overline{\tau_\alpha} + \delta \tau_{\alpha}$, where $\alpha = {\rm on, off, iso}$ and $\{\delta\tau_\alpha\}$ are three independent random variables with zero mean and variance $\overline{\delta \tau_\alpha^2} = \sigma_\alpha^2 \tau_\alpha^2$. For concreteness, we take a uniform distribution of $\delta\tau_\alpha$ and choose the same relative error for each stroke, $\sigma_\alpha = \sigma$.

In Fig.~\ref{fig:schedule_noise} we plot the results of a simulation of 100 random realisations of such a noisy protocol in the RL model. As one might expect, small timing errors lead to small changes in the overall dissipated work. Our results show that the resulting fluctuations in the dissipated work are of the same relative order as the error in the timing, $\sigma$. This holds true even for quite large relative errors $\sigma\sim 10$\% and relatively fast driving where $g^2\tau_{\rm tot}\sim 1$. For the parameters of Fig.~\ref{fig:schedule_noise}, the protocol with $k=2$ is approximately two times faster than the $k=1$ case. Hence, the SI remains advantageous since timing errors increase the dissipation by at most a few percent.

\section{Conclusions}\label{sec:conclusions}

We have put forward the idea of a speed-up to isothermality (SI), where an isothermal process is sped up by smoothly increasing (and decreasing) the system-bath interaction. This leads to faster isothermal processes while keeping the overall thermodynamic dissipation constant. As a consequence, our proposal allows for increasing the power of finite-time Carnot cycles \cite{Schmiedl2007,Schmiedl2007b,Esposito2010,CavinaSlow,Ma2018,abiuso2019optimal} and refrigerators \cite{Wang2012,Hu2013,Hernandez_2015,holubec2020maximum} without compromising their efficiency.

To obtain these results, we followed a two-fold approach. First, we  analytically constructed SI under two main assumptions:
\begin{enumerate}
\item Slow driving, allowing for an expansion of the dissipation as in Eq. \eqref{eq:WdissIII}. 
\item The timescale of thermalization satisfies $\tau_{\rm eq}\propto g^{-2}$, where $g$ quantifies the strength of the system-bath interaction, as expected in dissipative systems \cite{Breuer_Petruccione}. 
\end{enumerate}
Under these assumptions, we have shown that SI can decrease the time of a given isothermal process by several orders of magnitude, see Eq.~\eqref{eq:totalTime} and Fig.~\ref{fig:TvsKk}. This leads to faster decays of the dissipation with time (Sec. \ref{sec:optimalscaling}) and higher efficiencies at maximum power of finite-time Carnot engines and refrigerators (Sec.  \ref{sec:EMP}). 

Second, we have tested the analytically derived SI for two generic models of dissipation covering both bosonic and fermionic baths. In both  cases, we exactly simulated the system-bath evolution at strong coupling in order to account for all sources of dissipation. The obtained results confirm the  validity of our analytical considerations,  while strongly suggesting that SI can even be applied  beyond the weak coupling and slow driving regime. An analytical characterisation of SI in these regimes is an interesting and challenging future research direction. In this sense, it seems promising  to combine the ideas presented here with open systems techniques to deal with strong, time-dependent coupling such as the reaction-coordinate mapping \cite{Strasberg2016,Newman2017Performance,Strasberg2018,Restrepo2018,Nazir2018,newman2019quantum}, or more sophisticated tensor-network methods~\cite{Prior2010,Strathearn2018,Somoza2019,Brenes2019}. 

Another interesting direction is to characterise the work fluctuations due to such SI, which have been characterised in e.g. STA~\cite{Funo2017}, and other tradeoffs between thermodynamic cost and time enhancements~\cite{Campbell2017,Abah2019}. Indeed, because SIs allow for accessing larger energy scales, one expects that they shall generate higher work fluctuations~\cite{Funo2017}.  In this sense, we note that one expects a competing effect in the work fluctuations generated by a SI: because we are accessing stronger coupling and hence larger energy scales, one expects stronger fluctuations; however, for a fixed time, SI allow for decreasing dissipation, and in the quasistatic regime the minimisation of dissipation comes together with the minimisation of fluctuations, at least for commuting protocols~\cite{miller2019work}. A further interesting possibility is to combine these considerations with geometric optimal paths~\cite{Sivak2012,Zulkowski2015,Scandi,miller2019work}.

Quantum heat engines have been experimentally realised in a range of different architectures, including trapped ions~\cite{Rossnagel2016,Lindenfels2019}, nanomechanical resonators~\cite{Klaers2017}, nitrogen-vacancy centres~\cite{Klatzow2019} and quantum dots~\cite{Josefsson_2018}. Our theoretical results are applicable to platforms where the system-bath coupling can be tuned. We have discussed two specific possibilities --- ultracold atomic impurities and mesoscopic electronic devices --- but other setups may also be feasible. For example, reservoir engineering is possible in trapped-ion systems by controlling the vibrational degrees of freedom of the ions~\cite{Lemmer2018}. It is also worth stressing that the the proposed speed-ups are robust to imperfections in the control or timing of the driving, as enhancements are found for a large family of protocols. We have demonstrated this explicitly, showing that SI protocols remain advantageous even when the system-bath coupling strength or the timing of the strokes is noisy. Enhanced quantum heat engines via speed-ups to isothermality thus appear feasible with current or near-future technology. 

\hspace{5mm}

\emph{Acknowledgements.} We thank Ahsan Nazir for insightful discussions. N.P. acknowledges Exploring Quantum Matter and the Elite Network of Bavaria for financial support. M.S. acknowledges funding from the European Union's Horizon 2020 research and innovation programme under the Marie Sk\l{}odowska-Curie grant agreement No 713729, and from Spanish MINECO (QIBEQI FIS2016-80773-P, Severo Ochoa SEV-2015-0522), Fundacio Cellex, Generalitat de Catalunya (SGR 1381 and CERCA Programme). M.T.M. is funded by the  European Research Council (ERC) under the European Union's Horizon 2020 research and innovation program (Grant Agreement No. 758403). Some of our calculations were performed on the Lonsdale cluster maintained by the Trinity Centre for High Performance Computing. This cluster was funded through grants from Science Foundation Ireland.

\bibliography{library}

\appendix

\section{Extensions to more general dissipations} \label{sec:extensions}

In this section we consider the case in which the dissipation is of the form as in Eq.~\eqref{eq:WdissIII}:
\begin{align}
\label{eq:WdissRep}
W_{\rm diss}=\frac{1}{\tau}\, \int_{0}^{1}dt\,G_{\rho^{\rm th}_t}\hspace{-1mm}\left({\dot H_{t} , \dot{H}_{s}}\right)+\mathcal{O}\left(\frac{\tau^2}{\tau^2} \right),
\end{align}
where $G_{\rho^{\rm th}_t}$ is a bilinear form which depends only on the base point $\rho^{\rm th}_t$. This expression is generic, and it arises in the expansion of the entropy production rate $\dot \sigma_t$ in the quasi-static limit~\cite{Nulton1,Crooks,Scandi}:
\begin{align}
	\dot \sigma_t = G_{\rho^{\rm th}_t}\hspace{-1mm}\left({\dot H_{t} , \dot{H}_{s}}\right)+\mathcal{O}\left(\frac{\tau^2}{\tau^2} \right).
\end{align}
In particular, if the dynamics is described by the time dependent Liouvillian equation:
\begin{align}
	\rho = L_t[\rho],
\end{align}
where $L_t$ has for every $t$ only one thermal steady state and, moreover, the real part of all its eigenvalues is negative (this two conditions are sufficient to ensure thermalisation), then the integrand in Eq. \eqref{eq:WdissRep} is given at first order by~\cite{Scandi}:
\begin{align}
	G_{\rho^{\rm th}_t}\hspace{-1mm}\left({\dot H_{t} , \dot{H}_{s}}\right)  = -\beta\, \Tr\sqrbra{\dot H_{t} L^+_t  \hspace{-1mm}\left[ \mathbb{J}_{\rho^{\rm th}_{\beta}(H_s)} [\dot H_{t}] \right]},\label{entropyPrate}
\end{align}
where we defined the two operators:
\begin{align}
	\J_{\rho}[A]&:=\int_0^1 \hspace{-0.5mm}{\rm d}s \hspace{1mm}   \rho^{1-s}\left(A - \Tr[\rho A]\id\right)\rho^{s},\\
	L^+_t[A]&:=\int^\infty_0 \text{d}\nu \ e^{\nu L_t}\big(\rho^{\rm th}_\beta (H_s)\Tr\sqrbra{A}-A\big).\label{DrazinInverse}
\end{align}
Before going on, it should be noticed that the operator $\mathbb{J}_\omega$ is related to the generalised covariance through the equality:
\begin{align}
	{\rm cov}_{\rho} \left(  A, B \right) = \Tr\sqrbra{A\,\mathbb{J}_{\rho} [B] }.
\end{align}
Moreover, carrying out the integral in Eq. \eqref{DrazinInverse} in the eigenbasis of $L_t$ shows that the eigenvalues of $L^+$ are directly connected with the different thermalisation timescales in the system. In particular, in the case in which all the observables thermalises at the same rate, Eq. \eqref{entropyPrate} reduces to Eq. \eqref{eq:Wdiss}. 

Considering again the simplified case in which the derivative of the Hamiltonian is given by $\dot{H}=\dot{\lambda}_t X$, we have the chain of inequalities:
\begin{align}
|W_{\rm diss}|&=\frac{\beta}{\tau}\, \Bigg |\int_{0}^{1}dt\,\dot{\lambda}_t^2 \ \Tr\sqrbra{X L^+_t  \hspace{-1mm}\left[ \mathbb{J}_{\rho^{\rm th}_{\beta}(H_s)} [X] \right]}\Bigg|\leq\nonumber\\
&\leq \frac{\beta}{\tau}\, \sup_{t\in\sqrbra{0,1}}{\rm cov}_{\rho^{\rm th}_t} (X, X)  \int_{0}^{1}dt\,\dot{\lambda}_t^2 \tau^{\rm max}_{g(t)},
\end{align}
where we indicate with $\tau^{\rm max}_{g(t)}$ the biggest eigenvalue of $L^+_t$. Since during the turning on and off procedure we want to keep track of the dependence of the thermalisation timescale on the interaction strength, we keep this term inside the integral. This expression should be compared with Eq. \eqref{eq:boundWdis_tm} and \eqref{disOn} above.

As a final remark, the bound in Eq. \eqref{disOn} on the covariance can be improved to~\cite{Marshall1985}:
\begin{align}
\sup_{t\in\sqrbra{0,1}}{\rm cov}_{\rho^{\rm th}_t} (F, F)\leq 2\sup_{t\in\sqrbra{0,1}}\norbra{\average{V^{2}}_{\rho^{\rm th}_t}-\average{V}_{\rho^{\rm th}_t}^2}.
\end{align}
This quantity is expected to be finite even in the limit in which $||V|| \rightarrow \infty$.

\section{Solution of the resonant-level model}
\label{app:RLM}
In this appendix, we detail our approach to solve the resonant-level model described in  Sec.~\ref{sec:resonant_level_model}. Starting from the Hamiltonian given in Eqs.~\eqref{eq:H_S_fermion}--\eqref{eq:H_SB_fermion}, we derive the Heisenberg equations
\begin{align}
\label{Heisenberg_equation_a}
\dot{a}(t) & = -{\rm i} \varepsilon(t) a(t) - {\rm i} g(t)\sum_k \bar{\lambda}_k b_k(t),\\
\label{Heisenberg_equation_b}
\dot{b}_k(t) & = -{\rm i} \omega_k b_k(t) - {\rm i} g (t)\bar{\lambda}_k a(t).
\end{align}
The second equation can be formally solved to give 
\begin{equation}\label{bath_collective_solution}
\sum_k \bar{\lambda}_k b_k(t) = \xi(t) - {\rm i}\int_{t_0}^t {\rm d}t'\, \chi(t-t')g(t')a(t'),
\end{equation}
where we defined the noise operator 
\begin{equation}\label{noise_operator}
\xi(t) = \sum_k \bar{\lambda}_k {\rm e}^{-{\rm i}\omega_k(t-t_0)} b_k(t_0),
\end{equation}
whose Gaussian statistics with respect to the initial state define the memory kernel $ \chi(t-t') = \langle \{\xi(t),\xi^\dagger(t') \} \rangle $ and the noise correlation function $ \phi(t-t') = \langle [\xi(t),\xi^\dagger(t') ]\rangle $. These are given explicitly by
\begin{align}\label{memory_kernel_def}
\chi(t) & = \int\frac{\dd\omega}{2\pi}\, \ee^{-\ii\omega t} \bar{\mathfrak{J}}(\omega),\\\label{noise_kernel_def}
\phi(t) & = \int\frac{\dd\omega}{2\pi}\, \ee^{-\ii\omega t} \bar{\mathfrak{J}}(\omega)\tanh[\beta(\omega-\mu)/2],
\end{align}
where we defined a reduced (time-independent, dimensionless) spectral density $\bar{\mathfrak{J}}(\omega) =2\pi \sum_k \bar{\lambda}_k^2 \delta(\omega-\omega_k) = \mathfrak{J}(\omega)/g^2$. According to Eq.~\eqref{eq:spectral_density_fermion}, this is given by the top-hat function
\begin{equation}
    \bar{\mathfrak{J}}(\omega) = \Theta(\Lambda - |\omega|).
\end{equation}
Note that in Eq.~\eqref{noise_kernel_def}, for completeness, we allow for a finite chemical potential $\mu$. In the wide-band limit $\Lambda \to \infty$, the chemical potential can be set to zero without loss of generality by simply redefining all energies relative to $\mu$, which justifies our choice of $\mu=0$ in the main text. 

To obtain a tractable description, we approximate the memory kernel as
\begin{equation}\label{memory_kernel}
\chi(t) = \frac{\sin(\Lambda t)}{\pi t} \approx \delta(t).
\end{equation}
This is an exact equality (in the distributional sense) in the limit $\Lambda\to \infty$, and is a good approximation for finite $\Lambda$ so long as slowly varying functions and large times relative to the cut-off scale $\Lambda^{-1}$ are considered. The noise correlation function is approximated as 
\begin{align}
\label{noise_function}
\phi(t)  & = \int_{-\Lambda}^\Lambda\frac{\dd\omega}{2\pi}\, \ee^{-\ii\omega t} \tanh[\beta(\omega-\mu)/2] \notag \\
& \approx  \int_{-\infty}^\infty\frac{\dd\omega}{2\pi}\, \ee^{-\ii\omega t} \tanh[\beta(\omega-\mu)/2] \notag \\ &\quad  -  \int_{\Lambda}^\infty \frac{\dd\omega}{2\pi} \ee^{-\ii\omega t} +  \int_{-\infty}^{-\Lambda} \frac{\dd\omega}{2\pi} \ee^{-\ii\omega t} \notag \\
& = \frac{1}{\ii \beta}\left [ \frac{\ee^{-\ii\mu t}}{\sinh(\pi t/\beta)} - \frac{\cos(\Lambda t)}{\pi t/\beta}\right ].
\end{align}
On the second line, the integration domain is partitioned into three parts, and the approximation $\tanh(z) \approx \pm 1$ for $\pm z \gg 1$ is made. The first integral is essentially the Fourier transform of $\tanh(z)$, which is calculated by a standard contour integration, resulting in a geometric sum over Matsubara frequencies that evaluates to the first term in Eq.~\eqref{noise_function}. The remaining two integrals yield the second term in Eq.~\eqref{noise_function} with the help of the Sokhotski--Plemelj theorem. Note that this second term regulates the $1/t$ divergence as $t\to 0$ but is negligible (in the distributional sense) for time scales $t\gg\Lambda^{-1}$.  It can be shown that, within these approximations, the fluctuation-dissipation relation $\tilde{\phi}(\omega) = \tilde{\chi}(\omega)\tanh[\beta(\omega-\mu)/2]$ between the Fourier components of the memory kernel $\tilde{\chi}(\omega)$ and the noise spectrum $\tilde{\phi}(\omega)$ holds for all $|\omega|<\Lambda$. 

\begin{figure}
    \centering
    \includegraphics[width=\linewidth]{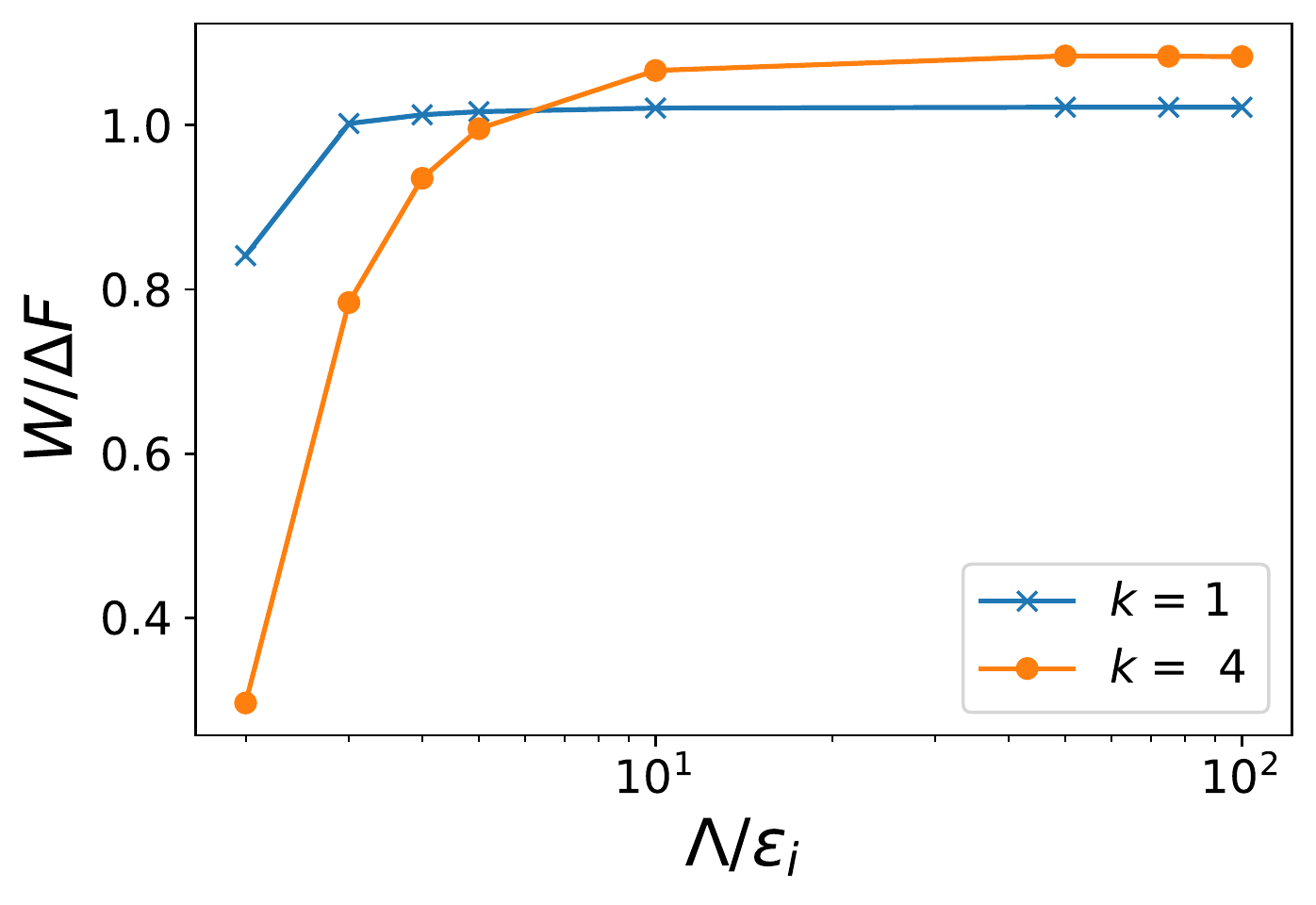}
    \caption{Convergence of the work done over an entire protocol, where $\tau_{\rm on} = k \tau_{\rm on}^{\rm weak}$ and $\tau_{\rm iso} = \tau_{\rm on}^{\rm weak}/k^2$, for various values of the cutoff $\Lambda$. Parameters: $\varepsilon_f = 2\varepsilon_i$, $g_0^2/\varepsilon_i = 0.1$, $\beta\varepsilon_i = 1$, $g_0^2\tau_{\rm on}^{\rm weak} = 2.5$ and $g_0^2\tau_{\rm iso}^{\rm weak} = 50$.}
    \label{fig:Lambda_convergence}
\end{figure}

As a consequence of Eq.~\eqref{memory_kernel}, Eq.~\eqref{Heisenberg_equation_a} reduces to a time-local differential equation
\begin{equation}\label{a_equation_closed}
\dot{a}(t) =  \left (-\ii \varepsilon(t) a(t)  - \frac{1}{2} g(t)^2\right )a(t) - \ii g(t)\xi(t),
\end{equation} 
which can be easily solved to find
\begin{equation}\label{a_solution}
a(t) = K(t,t_0) a(t_0) - \ii \int_{t_0}^t\dd t' \, K(t,t')g(t') \hat{\xi}(t'),
\end{equation}
where the propagator is given by Eq.~\eqref{eq:propagator}. Combining this with Eqs.~\eqref{bath_collective_solution}, \eqref{memory_kernel} and \eqref{noise_function}, and the fact that $\langle a^\dagger(t_0)b(t_0)\rangle  = 0$ for a factorized initial condition at $t_0 = 0$, we deduce Eqs.~\eqref{eq:level_occupation} and \eqref{eq:SB_correlation}. 

Our analysis relies on  two approximations, given by Eqs.~\eqref{memory_kernel} and \eqref{noise_function}. The former assumes that the dynamics is much slower than $\Lambda^{-1}$, while the latter requires that the temperature and chemical potential are much smaller than $\Lambda$. In particular, we require that $\varepsilon(t),  g(t)^2, \beta^{-1}, |\mu|$ and  $|\mu\pm \beta^{-1}|$ are all much smaller than $\Lambda$. For sufficiently large $\Lambda$, the work done over a complete isothermal protocol converges to a $\Lambda$-independent value, as we demonstrate in Fig.~\ref{fig:Lambda_convergence}. This confirms that our results are independent of the cutoff, which merely regulates the system-bath interaction energy that would otherwise diverge.

\clearpage 
\end{document}